\documentclass[AMSmath,usenatbib]{mn2e}
\usepackage{epsfig, bm}
\usepackage{aas_macros}

\def\s{{\rm\,s}}
\def\erg{{\rm\,erg}}

\def\eV{{\rm\,eV}}
\def\ang{{\rm\,\AA}}
\def\cm{{\rm\,cm}}

\def\km{{\rm\,km}}

\def\keV{{\rm\,keV}}

\def\g{{\rm\,g}}

\def\days{\rm\,days}

\def\K{{\rm\,K}}
\def\yr{{\rm\,yr}}

\def\rt{R_{\rm T}}
\def\rtcore{R_{\rm T, core}}
\def\rs{R_{\rm S}}

\def\mbh{M_{\rm BH}}
\def\msun{M_\odot}
\def\rsun{R_\odot}
\def\mstar{M_\star}

\def\rcore{R_{\rm core}}
\def\rapo{R_{\rm apo}}

\def\rph{R_{\rm ph}}

\def\redge{R_{\rm edge}}

\def\tfall{t_{\rm fallback}}

\def\mdotfb{\dot{M}_{\rm fallback}}
\def\mbhrat{\left(\frac{\mbh}{2\times 10^5\msun}\right)}

\def\rtcorerat{\left(\frac{\rtcore}{2\times 10^{11}\cm}\right)}
\def\rcorerat{\left(\frac{R_{\rm core}}{0.03\rsun}\right)}

\def\mcore{M_{\rm core}}
\def\mcorerat{\left(\frac{\mcore}{0.3\msun}\right)}

\def\nhepp{n_{\rm He^{++}}}

\def\teff{T_{\rm eff}}
\def\tpeak{t_{\rm peak}}

\def\LHeHa{L_{4686}/L_{\rm H\alpha}}
\def\lesssim{\mathrel{\hbox{\rlap{\hbox{\lower3.5pt\hbox{$\sim$}}}\hbox{$<$}}}}
\def\gtrsim{\mathrel{\hbox{\rlap{\hbox{\lower3.5pt\hbox{$\sim$}}}\hbox{$>$}}}}

\title[Tidal Disruption Candidate PS1-10jh]{Insights into Tidal Disruption of Stars from PS1-10jh}
\author[L. E. Strubbe and N. Murray]{Linda E. Strubbe$^{1,2}$\thanks{UBC Science Teaching and Learning Fellow. Email: linda@phas.ubc.ca}, Norman Murray$^{1}$ \\ 
$^{1}$Canadian Institute for Theoretical Astrophysics, 
University of Toronto, 60 St. George St., Toronto, ON, M5S 3H8, Canada\\
$^{2}$Department of Physics and Astronomy \& Carl Wieman Science Education Initiative,\\ 
University of British Columbia, 6224 Agricultural Rd., Vancouver, BC, V6T 1Z1, Canada\\}

\begin{document}
\date{Accepted . Received ; in original form }

\pagerange{\pageref{firstpage}--\pageref{lastpage}} \pubyear{2015}
\maketitle

\label{firstpage}
\begin{abstract}
Was PS1-10jh \citep{gezari12}, an optical/UV transient discovered by the Pan-STARRS Medium Deep Survey, the tidal disruption of a star by a massive black hole? We address two aspects of the problem: the composition of the putative disrupted object (using the spectroscopic data), and the energetics of the observed gas and radiation (using the photometric data). We perform photoionization calculations and compare with the observed lower limit of the line ratio $L_{\rm HeII\,\, 4686}/L_{\rm H\alpha}>5$ to argue that this event was {\it not} the disruption of a solar-type star, and instead was likely the disruption of a helium core \citep[as first proposed by][]{gezari12}. Disruption of such a dense object requires a relatively small central BH, $\mbh\lesssim 2\times 10^5\msun$.  We use the photometric data to infer that PS1-10jh comprised an outflow of $\sim 0.01\msun$ of gas, escaping from the BH at $\sim 1000\km\s^{-1}$, and we propose that this outflow was driven primarily by radiation pressure trapped by Thomson and resonance line scattering.  The large ratio of radiated energy to kinetic energy, $E_{\rm rad}/E_{\rm K}\sim 10^4$, together with the large value of $E_{\rm rad}\sim 2\times10^{51}\erg$, suggests that the outflow was shocked at large radius (perhaps similar to super-luminous supernovae or the internal shock model for gamma-ray bursts). We describe puzzles in the physics of PS1-10jh, and discuss how this event may help us understand future tidal disruptions and super-Eddington accretion events as well.

\end{abstract}


\section{Introduction}\label{intro}
On May 31, 2010, the Pan-STARRS Medium Deep Survey \citep{kaiser10} first detected a remarkable event called PS1-10jh \citep{gezari12}.  Over the next two months, the event rose in brightness by a factor of $\sim 200$, then declined over the subsequent year.  Follow-up observations showed colors and spectra inconsistent with a supernova (SN) or active galactic nucleus (AGN).  These authors interpreted the event as the tidal disruption of a helium-rich stellar core.

Tidal disruption events (TDEs) must happen every so often in the centers of galaxies, when a star passes so close to the central massive black hole (BH) that the BH's tidal gravity exceeds the binding gravity holding the star together.  The star disrupts, sending a fraction of stellar gas on elliptical orbits that eventually return to the BH; the stream of stellar gas shocks on itself and begins to accrete onto the BH, releasing a flare of electromagnetic energy that we in principle can observe \citep[e.g.,][]{lacy, rees}.  TDEs are interesting in their own right, and also as signposts of otherwise quiescent BHs, probes of stellar dynamics in the nuclei of galaxies, and informants about accretion physics.

A few dozen transient events have been identified as candidate TDEs over the last couple of decades \citep[e.g.,][]{komossa02,gezari09,bloom11}, but the advent of wide-field transient surveys is now dramatically improving the coverage of data available for any individual event \citep[e.g.,][]{gezari12,chornock14,arcavi14}.  PS1-10jh is special because:
\begin{enumerate}
\item It is the first event whose light curve follows the {\it rise} to peak brightness\footnote{PS1-10jh is also only the second optically-selected event to be followed up in real time, after PTF10iya \citep{cenko11}.} (not just the later-time decline),
\item It is the first event whose optical spectrum shows lines at all \citep[with the possible exception of TDE2;][]{vanvelzen11},
\item The spectrum shows He II emission but no hydrogen, suggesting that the event was the disruption of the core of a red giant star rather than the disruption of a main sequence star, and
\item As a consequence of (3), the host galaxy may harbor a particularly low-mass central BH.
\end{enumerate}
In addition, ultraviolet observations of PS1-10jh indicate a relatively cool effective temperature $\teff \sim 3\times 10^4\K$ that appears to have remained mysteriously constant for most of a year, while the luminosity ranged by a factor of $\sim 50$.

We seek to develop a model that explains the observations and offers insights into the tidal disruption process and accretion physics.

\subsection{Observed properties of PS1-10jh}
Figure 2 in \citet{gezari12} 
shows the observed light curve of PS1-10jh in optical ($g$, $r$, $i$, $z$ from Pan-STARRS) and ultraviolet (near-UV from the Galaxy Evolution Explorer: GALEX; \citealt{martin05}).  The event reached a maximum $g$-band luminosity of $2\times 10^{43}\erg\s^{-1}$ on July 12, 2010; 
 after peak, the event was detectable to Pan-STARRS for a little over a year (until August 2011), over which time the $g$-band luminosity faded by a factor of $\sim 50$.  GALEX observed the event about 20 days before $g$-band peak and again about 250 days after peak.
For the two epochs that GALEX data was available, \citet{gezari12} fit the optical and NUV points of the spectral energy distribution (SED): they find that the SED at {\it both} epochs can be fit by a blackbody of temperature $T = 2.9 \times 10^{4}\K$, assuming a galactic extinction of $E(B-V) = 0.013$ mag.  This temperature would imply a peak bolometric luminosity of $L_{\rm bol} \sim 2.2 \times 10^{44}\erg\s^{-1}$.  The extinction may actually have been greater than this, leading to a hotter blackbody temperature; the limit of $E(B-V) < 0.08$ mag from the He II line ratio (discussed below) implies a temperature of $5.5 \times 10^4\K$.

\citet{gezari12} followed up the event with an optical spectrum from MMT about 20 days before ($g$-band) peak and again about 250 days after peak. The optical spectra can be fit by stellar light from the host galaxy, plus a blackbody continuum from the event ($T=2.9\times 10^4\K$), and---intriguingly---broad He II emission lines.  In the spectrum at time $t_{\rm spect} = t_{\rm peak}-22\,{\rm days}$, two He II lines are visible, at rest wavelengths $\lambda = 4686\ang$ ($n = 4\rightarrow 3$) and $\lambda = 3203\ang$  ($n = 5\rightarrow 3$).  The line at $4686\ang$ has a linewidth of $9000\km\s^{-1}$ and luminosity of $L_{4686}=9\times 10^{40}\erg\s^{-1}$.  The line at $3203\ang$ is significantly fainter and its wavelength and linewidth are less constrained, but its presence is thought to confirm the identity of the $4686\ang$ line.  The ratio of line strengths limits the internal extinction to $E(B-V) < 0.08$ mag.  Surprisingly, there are {\it no} hydrogen or other emission lines detected in either epoch of spectroscopy.  The $4686\ang$ line has faded by a factor of 10 by the second epoch (at $t_{\rm peak}+254\,{\rm days}$), while the $3203\ang$ line is no longer visible then.

The location of the event was observed at 0.2 - 10 keV with Chandra 315 days after the peak in the $g$-band light curve; no source (with a spectral slope of $\Gamma = 2$, characteristic of AGN) was detected above the background to an upper limit of $L_X < 5.8\times 10^{41}\erg\s^{-1}$.  The event was followed up at 5 GHz by the Very Large Array in 2012 but not detected \citep{vanvelzen13}.

\subsection{Host galaxy and BH}
PS1-10jh was coincident with the center of a host galaxy at $z= 0.1696$ identified in the Sloan Digital Sky Survey \citep[SDSS;][]{SDSS} and UKIRT Infrared Deep Sky Survey \citep[UKIDSS;][]{ukidss}, whose
photometry is consistent with a stellar mass of $M_{\ast, \rm gal} = 3.6 \times 10^9 \msun$ \citep{gezari12}.  The spectrum of the host galaxy shows no narrow or broad emission lines (indicating a lack of star formation and AGN activity), and can be fit by a relatively old stellar population 1.4--5.0 Gyr old (suggesting PS1-10jh was not a core-collapse supernova).

The morphology of the galaxy is poorly constrained from the low-resolution images in SDSS and UKIDSS.  The low stellar mass inferred from the colors could suggest that the galaxy is a disk, while the lack of star formation and clear $4000\,{\rm \AA}$ break suggest that the host may instead be early-type.  
The Magorrian relation between bulge mass and BH mass allows us to make some estimates (though only poor) of the mass of the galaxy's central BH \citep{haring04}.  Assuming that the galaxy is purely a bulge of mass $M_{\ast, \rm gal} = 3.6 \times 10^9 \msun$, the mass of the galaxy's central BH would be expected to be $\mbh \sim 5 \times 10^6\msun$.  If the galaxy is instead predominantly a disk whose bulge comprises only $\sim 10\%$ of the total mass, the BH may be $\sim 10$ times less massive, $\mbh \sim 5 \times 10^5\msun$.  At the same time, the Magorrian relation is poorly constrained and shows high scatter at low masses, and so these estimates may not be so valid \citep[e.g.,][]{greene08,jiang11}.  In any case, the observations of the host galaxy are not inconsistent with a central BH of mass $\mbh \sim 10^5 - 10^6\msun$.

\subsection{Previous studies of PS1-10jh}
Because the spectrum of PS1-10jh showed He II emission lines but no hydrogen lines, \citet{gezari12} proposed that the event was the tidal disruption of the helium-rich core of a red giant that had previously been stripped of its hydrogen-rich envelope.  Various works since then have investigated the event and this interpretation further.  \citet{bogdanovic_cheng} studied how tidal heating of the core and orbital decay by emission of gravitational waves could lead to disruption of the core by a $\sim 10^6\msun$ BH; their work builds on previous work by \citet{macleod12} that investigated the partial disruption of red giant stars  (see further discussion of these papers in \S\ref{sec:discussion}). \citet{gezari12} considered in particular a red giant whose envelope had been stripped by a previous interaction with a massive BH \citep[as in][]{davies05}, and they approximated its mass and radius from observations of a red giant in a binary system whose envelope had been stripped by its companion \citep{maxted11}.
\citet{armijo13} studied PS1-10jh using a hydrodynamic code and estimated the accretion rate.

\citet{guillochon14} instead proposed that PS1-10jh was the disruption of a main-sequence star, appealing to simulations of AGN spectra where He II 4686 is brighter than H$\alpha$; these authors perform no photoionization calculations themselves, though \citet{gaskell14} do and state that their findings support \citet{guillochon14}'s idea.  They claim that ``over-ionization'' of hydrogen can suppress H$\alpha$ emission relative to He II 4686.  We investigate and discuss these ideas at length in Appendix \ref{sec:cloudy}, and explain that this idea is erroneous: instead, a high ionization fraction of hydrogen produces copious Balmer emission as ionized hydrogen recombines in steady state.  We thus conclude as \citet{gezari12} do, that PS1-10jh is likely the disruption of a helium-rich stellar core.  \citet{guillochon14} also suggest the presence of gas outside an accretion disk to reprocess and cool the radiation.

An investigation of the accretion physics and radiation processes that took place in PS1-10jh has so far been lacking.  \citet{guillochon14} do perform numerical simulations of disruption and fallback to the BH (though they disrupt a solar-type star rather than a helium core); however, their simulations lack radiation pressure (and accretion physics), which
we argue is critical for understanding PS1-10jh.

\subsection{Proposed model and plan of the paper}

In this work,
we separately draw inferences from the continuum photometric observations, and from the spectroscopic data, and bring these together to propose a physical model for the event.

In \S\ref{sec:contobs}, we use photometric observations to infer the presence of an outflow and estimate its velocity, mass, and geometry; then in \S\ref{sec:kin_ene} we build on the photometric inferences to study the kinematics and energetics of the outflow.  In \S\ref{sec:heII_emission}, we study the observed He II emission lines to argue that the lack of observed hydrogen emission essentially rules out the possibility that PS1-10jh was the disruption of a solar-type star, and suggest the presence of a resonance-line-driven wind. \S\ref{sec:hecore} investigates the possibility that PS1-10jh was the disruption of a helium core, and \S\ref{sec:interpretation} describes our proposed model (summarized in the following paragraph). We compare the properties of PS1-10jh with those of supernovae in  \S\ref{sec:SNe}, and wrap up with a discussion in \S\ref{sec:discussion}.  The Appendices offer detailed investigation of photoionization processes in PS1-10jh.

We argue that PS1-10jh was the disruption of the helium core of a red giant star, and propose the following model.
 Only hours following disruption, stellar gas completed elliptical orbits around the BH. The gas collided with itself and shocked, forming a highly optically thick region around the BH, and producing huge quantities of photons that were trapped by electron scattering.  The majority of the gas and radiation fell directly into the BH, but a small amount of gas ($\sim 0.01\msun$) escaped at a surprisingly slow velocity ($\sim 1000\km\s^{-1}$), likely driven by radiation pressure.  Very little of that radiation was able to escape until the expanding gas became optically thin---by which time the radiation had already lost most of its energy to adiabatic expansion. We suggest that the bulk of the observed radiation instead had a different source: the expanding gas experienced another shock at larger radius that converted expansion kinetic energy back to photons (perhaps similar to a superluminous supernova or internal shock in a gamma-ray burst; e.g., \citealt{gal-yam12,grb_internal}). Meanwhile, in the outer edge of the flow, radiation pressure trapped by resonance lines accelerated a wisp more gas outward from the photosphere at a somewhat faster velocity (i.e., a line-driven wind). The helium recombination cascade in this wind was observed as broad He II lines.

\section{Inferences from the continuum observations, Part I}
\label{sec:contobs}

\subsection{Radius and velocity of the photosphere}\label{sec:rad_photo}

We can use the optical light curve and the temperature of the photosphere $\teff$ to estimate the radius of the photosphere of the flow $\rph(t)$ \citep[see also][]{chornock14, arcavi14}.  We begin by considering $\teff$, approximately constrained to be $3 \times 10^4\K \lesssim \teff \lesssim 5.5\times 10^4\K$, as we explain below.

\citet{gezari12} use the optical and near-UV spectral energy distribution at $\tpeak-18\days$ and $\tpeak+245\days$ (where $t_{\rm peak}$ is defined as the time of peak emission in $g$-band) to infer a blackbody temperature of $\teff = 2.9\times 10^4\K$ at both epochs, assuming a standard extinction law $E(B-V)=0.013$ mag; if the extinction is greater than this, the temperature would be larger.  These authors derive a second temperature constraint from the first epoch of spectroscopic data ($\tpeak-22\days$):  at this time, the line ratio of the two observed He II lines indicates that the extinction is $E(B-V)<0.08$ mag, implying that the temperature is $\teff < 5.5\times 10^4\K$.  \citet{gezari12} also find that the late-time non-detection with Chandra at $t_{\rm peak}+315\days$ indicates an upper limit of $\teff \lesssim 2.5\times 10^5\K$ for $L_{\rm bol}\lesssim 10^{44}\erg\s^{-1}$.
We add an additional constraint that $\teff \gtrsim 3\times 10^4\K$ from the presence of doubly ionized helium (He$^{++}$) at both epochs of spectroscopic data ($\tpeak-22\days$ and $\tpeak+254\days$) (see \S\ref{sec:he_equilib}).  For simplicity, we assume that the temperature is constant in time, and we make calculations for two fiducial temperatures $T_{\rm eff} = 3\times 10^4\K$ and $5.5\times 10^4\K$.

\begin{figure}
\centerline{\epsfig{file=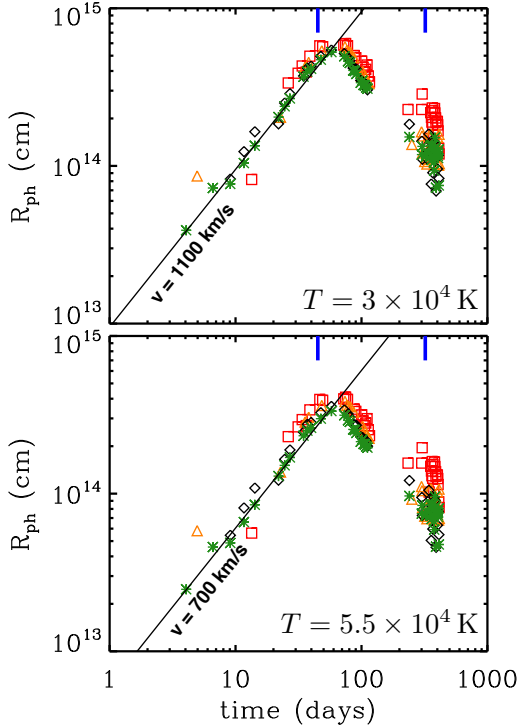, width=3in}}
\vspace{-0.4cm}
\caption{Radius of the photosphere calculated from equation (\ref{eq:R(t)}) using the different optical photometric datapoints and assuming a constant blackbody temperature $T_{\rm eff} = 3\times 10^4\K$ (top) and $5.5\times 10^4\K$ (bottom).  Green asterisks represent $g$-band data, black diamonds $r$-band, orange triangles $i$-band, and red squares $z$-band.  The black line in each panel represents expansion at a constant velocity, $R = v_0 t$ with $v_0 = 1100\km\s^{-1}$ (top) and $v_0 = 700\km\s^{-1}$ (bottom); the beginning of the expansion ($t=0$) is taken to be 67 days before peak luminosity. The blue tick marks denote the times when a spectrum was taken.  \label{fig:roft}}
\end{figure}

We now estimate the radius of the photosphere,
\begin{equation}\label{eq:R(t)}
R_{\rm ph}(t) = \left(\frac{\nu L_\nu(t)}{4\pi^2\nu B_\nu(\teff(t))}\right)^{1/2} \, ,
\end{equation}
where $\nu L_\nu$ is the luminosity in a given waveband of frequency $\nu$, and $\nu B_\nu$ is the blackbody function.  In Figure \ref{fig:roft}, we plot $R_{\rm ph}(t)$ using the $g$, $r$, $i$, and $z$-band data from \citet{gezari12}. Time is plotted starting from an assumed zeropoint at $t_{\rm peak}- 67\days$, which is about 4 days before the first detection\footnote{The cadence of the Medium Deep Survey is about 3 days \citep{kaiser10}, and later detections take place at spacings of $\sim 3 - 12$ days, so it is reasonable that the beginning of the event  was 4 days before the first detection.  If the event began slightly earlier, the growth of the photosphere would be slightly slower than linear with time; if it began slightly later, the growth would be slightly faster than linear with time.}. For a given temperature, calculations from the different optical bands give similar values of $R_{\rm ph}(t)$ (as expected if the dominant opacity is produced by a wavelength-independent process like Thomson scattering). A lower assumed temperature of course gives larger values of $R_{\rm ph}$, from $R_{\rm ph}(t_{\rm peak})\sim 6\times 10^{14}\cm$ for $3\times 10^{4}\K$, down to $R_{\rm ph}(t_{\rm peak})\sim 4\times 10^{14}\cm$ for $5.5\times 10^{4}\K$.

Overplotted are curves depicting expansion at a constant velocity $R = v_0t$ for two values of $v_0$ close to $1000\km\s^{-1}$.
We note that this expansion velocity is similar to the escape velocity at $\rph(t_{\rm peak})$ from a massive BH,
\begin{equation}\label{eq:vesc}
v_{\rm esc}(r) \sim 3000 \left(\frac{\mbh}{2\times 10^5\msun}\right)^{1/2}\left(\frac{6\times 10^{14}\cm}{r}\right)^{1/2}\km\s^{-1} \, ;
\end{equation}
it is similar to but somewhat lower than the half linewidth\footnote{See \S\ref{sec:linedriven} for discussion of the half versus full linewidth.} of the He II $4686\ang$ line, $4500\km\s^{-1}$ \citep{gezari12}.

Note that this analysis assumes that the observed emission is blackbody, although the dominant opacity is due to Thomson scattering rather than true absorption.  This effect may increase the size of the photosphere by a factor of a few above our estimates in this section.  Please see Appendix \ref{sec:scat_flow} for further discussion.

\subsection{Geometry}
\label{sec:geometry}
\begin{figure}
\centerline{\epsfig{file=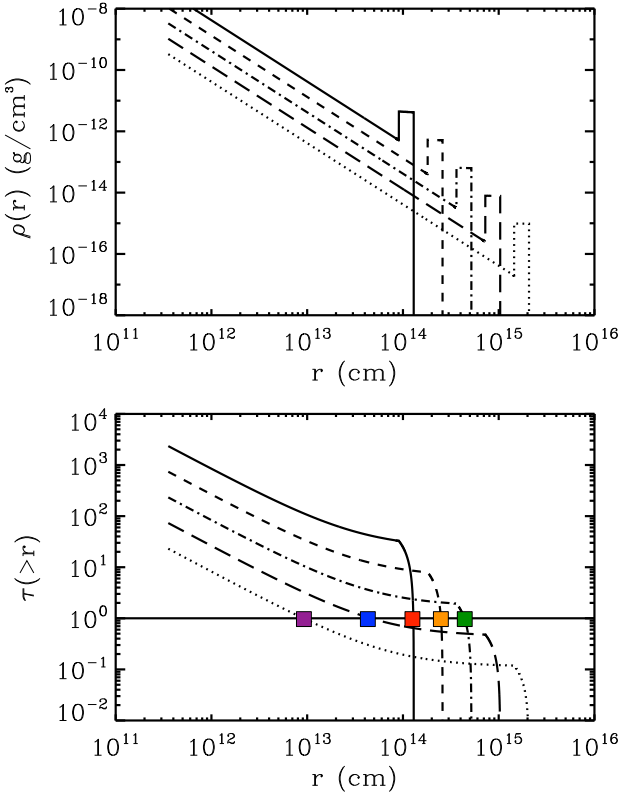, width=3.5in}}
\vspace{-0.4cm}
\caption{{\bf Upper panel}:  Schematic of our proposed density profile for the outflow from PS1-10jh.  Curves from top to bottom represent density profiles at times spaced by factors of 2.  The density profile is composed of a lower-density power-law inside a denser shell of gas of fixed mass (whose width has been enhanced for clarity).  {\bf Lower panel}:  Optical depth profiles corresponding to the density profiles in the upper panel, showing the optical depth from a given radius $r$ to infinity.  A horizontal line designates the location of the photosphere where $\tau(>\rph) \sim 1$.  The colored squares represent the location of the photosphere as a function of time, from red to orange to green to blue to purple. First the photosphere is confined to the edge of the shell and the photosphere moves out with time (red, orange); then the shell becomes optically thin (green) and the photosphere moves in with time (blue, purple).  \label{fig:rhotaushell}}
\end{figure}

The area, and hence the inferred radius, of the photosphere is clearly seen to rise and then fall (as a consequence of the rising then falling light curve, at constant temperature).  We propose that the flow comprises an expanding narrow shell of dense gas enclosing a lower-density region inside.  The shell has mass $M_{\rm shell}$, radius $R_{\rm shell}(t)$, width $\Delta R(t)$, and density $\rho_{\rm shell}(t)$.  We plot our proposed density profile schematically in the upper panel of Figure \ref{fig:rhotaushell}.  Early on, while we see the photosphere grow, the shell is optically thick, and so the photosphere lies at the outer edge of the shell ($\rph \sim R_{\rm shell}$): the photosphere is growing with the expanding shell.  (\citealt{sq09} referred to this situation as ``edge-dominated.'')  This is shown by red, orange, and green squares in the lower panel.  Later on, expansion causes the density in the shell to fall sufficiently that it becomes optically thin, and so we begin to see the region inside it; the photosphere then moves inward with time through this inner region (blue and purple squares in the lower panel of the figure).  At the peak of the light curve, the optical depth across the shell is unity. Since there is no sharp feature in the light curve at the time of maximum luminosity, we infer that any drop in the density at the inner edge of the shell is moderate.

We can use this model to estimate the mass and density in the shell, using its opacity.  In the likely density\footnote{For temperatures $T \sim 10^4 - 10^5\K$, electron scattering is generally the dominant opacity source for number densities up to at least $n \sim 10^{12}\cm^{-3}$.  At the earliest times, when the density in the shell was perhaps $10^3$ times its value at $t_{\rm peak}$, other sources of opacity may have contributed as well, but still only slightly.} and temperature range for the flow, the dominant source of opacity is electron scattering \citep{opal}.  Assuming the gas is composed primarily of doubly ionized helium (see \S\ref{sec:heII_emission}), this opacity is $\kappa_{\rm es,He} = \sigma_{\rm T}/2m_{\rm p} = 0.2\cm^2\g^{-1}$
(because there are two electrons for every doubly-ionized helium nucleus, and each ion weighs $4m_{\rm p}$).  

At the peak of the light curve $t_{\rm peak}$, we have $\tau=\kappa_{\rm es,He}\Sigma_{\rm shell}\sim 1$, where $\Sigma_{\rm shell}\equiv M_{\rm shell}/(4\pi R_{\rm shell}^2)$ is the surface density (assuming spherical geometry) and $R_{\rm shell} \sim \rph$.  
The mass is then
\begin{equation}\label{eq:mshell}
M_{\rm shell} \sim \frac{4\pi R_{\rm ph}^2}{\kappa_{\rm es,He}} 
\sim 0.01\msun \left(\frac{R_{\rm ph}}{6\times 10^{14}\cm}\right)^2 ,
\end{equation}
about $\sim 10\%$ of (half) the mass of a helium core (see \S\ref{sec:energetics} and \S\ref{sec:rt_MBH}).

We infer that the density in the shell is
\begin{eqnarray}\label{eq:rhoshell}
\rho_{\rm shell}(t_{\rm peak}) & \sim & \frac{1}{\kappa_{\rm es,He} \Delta R} \\
& \sim & 8\times 10^{-14} \left(\frac{0.1}{\Delta R/R_{\rm ph}}\right)^{-1}\left(\frac{R_{\rm ph}}{6\times 10^{14}\cm}\right)^{-1}\g\cm^{-3} \, , \nonumber
\end{eqnarray}
corresponding to a number density $\nhepp \sim 1 \times 10^{10}\cm^{-3}$. 

It will be useful later to have an estimate of the number density of helium\footnote{In \S\ref{sec:cloudy}, we test ideas about the composition of the gas by assuming that the gas has cosmic abundances, rather than being pure helium. In this scenario, where the bulk of the gas is hydrogen, the mass density in the shell would be half, and the hydrogen number density would be twice, the values we infer assuming pure helium: i.e., $8\times 10^{10}\cm^{-3}$ or $1 \times 10^{11}\cm^{-3}$, for $\teff = 3\times 10^4\K$ or $\teff = 5.5 \times 10^4\K$. \label{ft:nspect}} ions in the shell, $n_{\rm spect}$, at the time that the first spectrum was taken, $t_{\rm spect} = t_{\rm peak}- 22\days$.  In our picture, the density in the expanding shell decreases with time as $n_{\rm shell} \propto R_{\rm shell}^{-3} \propto t^{-3}$ (assuming that the ratio $\Delta R / R_{\rm shell}$ stays fixed at $\sim 0.1$).  This gives
\begin{equation}\label{eq:nspect}
n_{\rm spect} \sim 4\times 10^{10}\cm^{-3}
\end{equation}
(which we'll take as our fiducial value) for $\teff = 3\times 10^4\K$, or $n_{\rm spect} \sim 6\times 10^{10}\cm^{-3}$ for $\teff = 5.5 \times 10^4\K$.

\section{Inferences from the continuum observations, Part II}
\label{sec:kin_ene}

\subsection{Kinematics}
\begin{figure}
\centerline{\epsfig{file=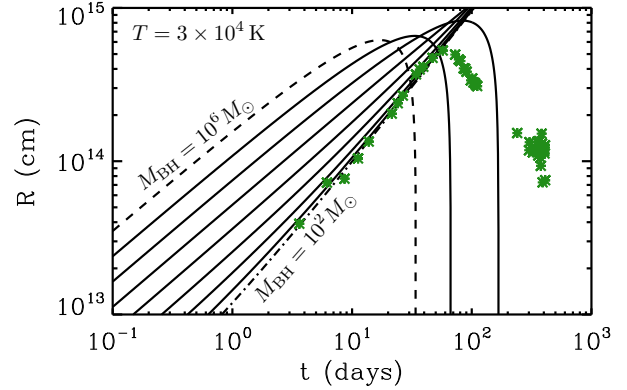, width=3.5in}}
\caption{Black curves: ballistic trajectories of particles in the gravitational potential of BHs of varying mass, assuming no acceleration due to radiation pressure.  Trajectories are fixed to have velocity $1100\km\s^{-1}$ at radius $6\times 10^{14}\cm$, to match our results for $\rph$ at peak for $T = 3\times 10^4\K$ (Figure \ref{fig:roft}).  Dashed curve is for $\mbh = 10^6\msun$, dot-dashed curve is for $\mbh=10^2\msun$, and curves are logarithmically spaced by $10^{1/2}$ in between.  Green asterisks represent $\rph$ (again from Figure \ref{fig:roft}).  Best agreement is for the smallest BH masses, indicating that $\mbh$ was small ($\lesssim 10^3\msun$) or that radiation pressure was important (for weakening the effective gravity of the BH on the particles). The observed recession of the photosphere does not at all match the shape of the fall exhibited by bound trajectories ($\mbh \gtrsim 10^5\msun$).  We propose that the photosphere appears to recede because the gas density decreases, rather than corresponding to physical motion of particles inwards.    \label{fig:kinematics}}
\end{figure}

We have shown that during the light curve's rise, the growth of the photospheric radius is close to linear in time (with a velocity $v_0 \sim 1000\km\s^{-1}$), and we hypothesized that the expanding photosphere is coincident with the edge of an expanding shell of gas. In Figure \ref{fig:kinematics} we study the kinematics of gas particles in the gravitational potential of BHs of varying mass, assuming ballistic motion.  We overplot our results for $\rph$ from Figure \ref{fig:roft} assuming $T = 3\times 10^4\K$, and fix the trajectories to match $v_0$ at peak radius $\rph$.

The ballistic trajectories are close to the data for low BH mass, where the BH's gravity has little effect on the particles' motion.  This suggests for PS1-10jh either that $\mbh$ is small ($\lesssim 10^3\msun$) or that radiation pressure was important (for reducing the effective gravity of the BH on the particles).  However, it would be surprising to find such a small central BH in this reasonable-sized host galaxy \citep[e.g.,][]{haring04}; moreover, the tidal disruption of a star by such a small BH would lead to incredibly super-Eddington feeding rates (see equation \ref{eq:mdotomdotedd}), making radiation pressure even more important.  For these reasons, we argue that this figure indicates that radiation pressure is important for the motion of gas particles in PS1-10jh.  

The observed gradual recession of the photosphere after 67 days has a very different shape from that of the plummeting fall exhibited by bound trajectories ($\mbh \gtrsim 10^5\msun$).  We conclude that the photosphere does not recede because of ballistic inward motion.  Rather, we propose that after the shell becomes optically thin at $t_{\rm peak}$, we are able to see through it outflowing gas with a run of density $\rho(r)$ that allows us to see deeper (in the Euclidian sense) towards the BH as time increases.

\subsection{Energetics}
\label{sec:energetics}
We can construct a picture of the energy flow in PS1-10jh by estimating energy scales for PS1-10jh:  radiation\footnote{The radiation energy is much greater than the thermal energy of the gas, so we neglect the latter.}, kinetic, and potential.  \citet{gezari12} assume a constant temperature of $3\times 10^{4}\K$ to estimate
the energy in radiation that escaped from the flow,
\begin{equation}\label{eq:Eradbolout}
E_{\rm rad, bol, out} = \int L_{\rm bol}(t)\, dt \sim 2 \times 10^{51}\erg \, .
\end{equation}
For comparison, we perform a similar calculation for the total energy emitted only in the observed bands (making no assumption about temperature), and find that this quantity is about 3 times smaller,
\begin{equation}
E_{\rm rad, obs, out} = \int L_{\rm obs}(t)\, dt \sim 6 \times 10^{50}\erg \, ;
\end{equation}
since this is a firm lower limit, and in any case these quantities are the same order of magnitude, we take $E_{\rm rad, out} \sim 2 \times 10^{51}\erg$ to be at least approximately correct.

We discuss how much radiation energy may have been present in the gas at early times in \S\ref{sec:adiabatic_cooling}.
The kinetic energy of the expanding shell of gas is
\begin{eqnarray}
E_{\rm K} & = & \frac{1}{2}M_{\rm shell} v_0^2 \nonumber \\
& \sim & 1 \times 10^{47} \left(\frac{M_{\rm shell}}{0.01 \msun}\right) \left(\frac{v_0}{1000\km\s^{-1}}\right)^2 \erg \, .
\end{eqnarray}
The kinetic energy is four orders of magnitude less than the radiation energy, $E_{\rm K}\sim 10^{-4}E_{\rm rad,bol,out}$, very different from supernovae, which have $E_{\rm K}\sim 10^2E_{\rm rad,bol,out}$ \citep{2009ARA&A..47...63S}, or even superluminous supernovae, which have $E_{\rm K}\sim E_{\rm rad,bol,out}$. We discuss this further in \S\ref{sec:adiabatic_cooling} and \S\ref{sec:SNe}. We note that $E_{\rm K}\propto R_{\rm ph}^4$, so that the inferred peak photospheric radius would have to be ten times larger than our estimate for the energy ratio of PS1-10jh to match that of superluminous supernovae.

If we now assume that the gas was launched from the innermost stable circular orbit (ISCO) around a non-spinning BH (at $R_{\rm launch} \sim 3\rs$, with $\rs = 2G\mbh/c^2$ the Schwarzschild radius: see eq. \ref{eq:rtcore}), we infer that by the time the gas was far from the BH, it had gained potential energy
\begin{eqnarray}\label{eq:Ep}
\Delta E_{\rm P} & \sim & \frac{G\mbh M_{\rm shell}}{R_{\rm launch}} \nonumber \\
& \sim & 3\times 10^{51} \left(\frac{M_{\rm shell}}{0.01 \msun}\right)\left(\frac{R_{\rm launch}}{3R_{\rm S}} \right)^{-1} \erg \, ,
\end{eqnarray}
similar in magnitude to the radiation energy and far exceeding the kinetic energy.  The total energy observed from PS1-10jh is thus
\begin{eqnarray}
E_{\rm tot,out} & \sim & E_{\rm rad,bol,out}+E_{\rm K}+\Delta E_{\rm P} \nonumber \\
& \sim & 5 \times 10^{51}\erg \, ,
\label{eq:Etot}
\end{eqnarray}
assuming our fiducial parameters.

Finally, supposing that PS1-10jh was indeed the disruption of a star, we estimate the energy released by the accretion of (half of) a stellar mass of gas onto a BH,
\begin{eqnarray}\label{eq:Edis}
E_{\rm dis} & \sim & \frac{\eta M_\ast c^2}{2} \\
& \sim & 3 \times 10^{52}\left(\frac{\eta}{0.1}\right)\left(\frac{M_\ast}{0.3\msun}\right)\erg \, ,
\end{eqnarray}
where the factor of 1/2 accounts for only the bound half of the disrupted star, $\eta \sim 0.1$ is a typical efficiency of converting accretion into radiation, and we have normalized to the mass of the helium core of a solar-mass red giant star (see \S\ref{sec:rt_MBH}).

Thus we uncover a hierarchy of energy scales,
\begin{equation}
E_{\rm K} \ll E_{\rm rad,bol,out} \sim \Delta E_{\rm P} \ll E_{\rm dis} \, .
\end{equation}
These energy scales offer insights and puzzles for our understanding of PS1-10jh, which we describe in \S\ref{sec:interp_energy}.

\section{Inferences from the He II observations}\label{sec:heII_emission}

\subsection{Not a main-sequence stellar disruption}
As described in the introduction, the optical spectrum of PS1-10jh shows the emission lines He II 4686$\ang$ and 3203$\ang$, but no hydrogen lines---in particular, no H$\alpha$ or H$\beta$.  \citet{gezari12} interpreted these observations to mean that PS1-10jh was the disruption of a helium-rich object, such as the stripped core of a red giant star.  \citet{guillochon14} and \citet{gaskell14} instead propose that PS1-10jh was the disruption of a main-sequence star, based on photoionization calculations for AGN (using the publicly available code Cloudy: \citealt{cloudy13}); they argue that the observed line ratio lower limit of He II 4686 to Balmer H$\alpha$ can occur for photoionized gas of cosmic abundances, and does not require an enhanced abundance of helium.

In Appendix \ref{sec:cloudy}, we extensively investigate these claims by performing and analyzing our own Cloudy photoionization calculations. Here we briefly outline the important physical processes and then discuss results.  The gas where the lines are produced has high enough density to be easily in photoionization equilibrium:  every photoionization is quickly balanced by a recombination, in which the ion joins an electron, and the electron ``cascades'' down to the ground state, releasing line-transition photons as it goes.  This mechanism by far dominates line emission of He II 4686 and H$\alpha$; the idea that the hydrogen could be ``over-ionized'' and thus produce {\it less} emission is incorrect---more ionized gas produces more recombination events and so {\it more} line emission.

One way to suppress H$\alpha$ relative to He II 4686 is to have a significant population of {\it neutral} hydrogen that absorbs H$\alpha$ photons---but few enough (singly-ionized) helium ions that He II 4686 is not too absorbed as well.  Relatively high gas densities and large incident fluxes generally give rise to the most relevant conditions.

However, our Cloudy calculations uncover significant issues, especially:
\begin{enumerate}
\item The calculations referenced and presented by \citet{guillochon14} and \citet{gaskell14} are for stationary gas, but the broad linewidth of He II 4686 indicates large velocity gradients, which would diminish optical depth significantly:  H$\alpha$ would not be effectively absorbed and suppressed;
\item Even if we make the same assumptions as these authors, we find that the ratio of He II 4686 to H$\alpha$ doesn't reach the observed lower limit for any density and incident flux level, especially including the appropriate parameters inferred from the continuum observations (\S\ref{sec:contobs});
\item Most problematically, Cloudy has serious convergence issues in much of the region of parameter space where these authors claim PS1-10jh would lie (high density and large incident flux), and so the numerical results likely are not trustworthy anyway (R. Porter, personal communication). We discuss this further in Appendix \ref{sec:cloudy}.
\end{enumerate} 

In view of these results, we argue that it is highly unlikely that the line-emitting gas of PS1-10jh had the cosmic abundances of hydrogen and helium: thus it was probably {\it not} the disruption of a solar-type star.  Instead we argue for the disruption of a helium-rich object such as the stripped core of a red giant, as suggested by \citet{gezari12}.

\subsection{Origin of the He II $4686\ang$ line}
\label{sec:emission_measure}
The He II $4686{\rm \AA}$ line is produced through the recombination cascade ($n=4$ to $n=3$) when He$^{++}$ recombines with an electron to produce He$^+$.  Provided a region of helium gas is (1) predominantly doubly ionized, (2) has had time to reach ionization equilibrium, and (3) is optically thin to the $4686\ang$ photons (see \S\ref{app:heII_lines} for justification of these three assumptions), the line luminosity will be
\begin{equation}\label{L4686}
L_{4686} \sim (h \nu_{4686})\alpha_{4686} (EM) \, ,
\end{equation}
where $h \nu_{4686} = 4.24\times 10^{-12}\erg$ is the energy of the transition, 
$\alpha_{4686} = 8.20 \times 10^{-14}\cm^3\s^{-1}$ is the effective Case B recombination coefficient\footnote{We make calculations using this temperature which is close to the temperature inferred for the photosphere $T\gtrsim 3\times 10^4\K$. Although the gas outside the Thomson photosphere at $\rph$ may not have exactly this temperature (since the gas is optically thin in the continuum), its temperature likely is not too different from this since the gas remains coupled to the radiation through line transitions.  The recombination coefficient scales roughly as $T^{-0.6}$ \citep{vernerferland}.} at $T=4\times 10^4\K$ \citep[p. 81]{osterbrock}, and $(EM) \equiv n_{\rm e}n_{\rm He^{++}} V$ is the emission measure of the line, with
$n_{\rm e}$ and $n_{\rm He^{++}}$ the number densities of free electrons and He$^{++}$, and $V$ the volume of the emitting region.  Using the  luminosity $L_{4686} = 9\times 10^{40}\erg\s^{-1}$ observed at $t_{\rm peak}-22\days$, we infer that the emission measure of the gas producing the He II $4686\ang$ line in PS1-10jh is
\begin{equation}
(EM) \sim 3 \times 10^{65}\cm^{-3} \, .
\end{equation}

We now compare this observed $(EM)$ with the emission measure of the region just outside the Thomson photosphere, $\rph$.  We make the assumption that the radial thickness of the line-emitting region is similar to $\rph$, because of the observation that the He II lines are centered on the galaxy's lines (if the emitting region instead were thin, most of the receding part of the flow would be blocked by the Thomson photosphere); the volume of the region is thus
\begin{equation}
V \sim ({\rm few}\times\rph)^3 \sim 10^{45}\cm^3 \, ,
\end{equation}
where we have used $\rph(t_{\rm spect}) \sim 4\times 10^{14}\cm$ at $t_{\rm spect} = t_{\rm peak}-22\days$.  If the density in the gas is similar to the density we infer for the shell at that time ($\nhepp \sim 4\times 10^{10}\cm^{-3}$; see eq. \ref{eq:nspect}), the emission measure in that gas would be 
\begin{equation}
(EM) \sim 10^{66}\cm^{-3} \, ,
\end{equation}
quite similar to the emission measure inferred from the line luminosity.  This lends additional support to our interpretation that the line-emitting region lies just outside the electron-scattering photosphere.

\subsection{Line-driven wind and ultraviolet absorption lines}
\label{sec:linedriven}

Broad-absorption-line quasars (BAL QSOs) and O-stars both show evidence for line-driven winds, in which gas at the outskirts of these objects absorbs luminous ultraviolet radiation via resonance line transitions (from ground state to higher energy level), and is accelerated outward by radiation pressure.  Observational evidence for the presence of these winds is provided by spectra showing blueshifted resonant absorption lines. The spectral energy distributions of O-stars are essentially those of blackbodies with $T_{\rm eff}\sim 20,000-50,000{\rm K}$; those of BAL QSOs show a similar blackbody-like UV bump, supplemented by strong X-ray emission. By analogy to these objects, we expect UV-bright PS1-10jh to have produced a line-driven wind as well.  We make some estimates about this wind and offer some predictions for observations of similar events in the future.

In \S\ref{sec:rad_photo}, we inferred the presence of a shell, whose expansion (at $v_{\rm shell}\sim 1000\km\s^{-1}$) we suggest is driven by radiation pressure from photons trapped by Thomson scattering. In the outer reaches of this shell, beyond $\rph$, the gas is optically thin to Thomson scattering. This outer gas should be subject to the same line-driven instability that drives high-velocity winds from O-stars. Since the observed spectrum shows us that He$^{+}$ was present in PS1-10jh, singly ionized helium is likely to provide a substantial line force for driving the wind; in \S\ref{sec:tau_heII}, we show that the He Lyman $\alpha$ transition at $E_{\rm He L\alpha} = 40.8\eV$ is highly optically thick.  By analogy with O-stars, we expect that there are many resonance metal lines that contribute to driving the wind; the resultant accelerations ensure a large radial velocity gradient in the gas, broadening the lines substantially.  In \S\ref{sec:mdotwind}, we speculate about the mass-loss rate of the wind.

Line-driven winds from O-stars have terminal velocities somewhat larger than the escape velocity at the wind's launch radius, typically by a factor of $3$ \citep{castor75,abbott82,howarth89,lamers99}:  radiation pressure gives the gas order unity of its gravitational binding energy.  The same is believed to be true for BAL QSO winds \citep{murray95,proga00}. In PS1-10jh, the observed half-width of the He II 4686 line corresponds to $v_{\rm line\, wind}\sim 4500\km\s^{-1}$, which we infer to be the terminal velocity of a wind.
This is a factor of a few times faster than the inferred expansion velocity of the continuum-driven shell.  We thus suggest a picture in which the outer layers of gas in the shell (just beyond $\rph$) are additionally accelerated by radiation pressure on ultraviolet resonance lines to velocities somewhat faster than the shell's velocity.  The observed He II emission lines are likely produced in the outer parts of the shell (as suggested in \S\ref{sec:emission_measure}) and in the faster (lower-density) line-driven wind beyond. The observed velocity of the line-emitting gas together with the size of the launch radius ($\sim \rph$) offers another estimate of the mass of the BH,
\begin{eqnarray}\label{eq:vesc_MBH}
\mbh & \sim & \frac{v_{\rm line\, wind}^2\rph}{2G} \nonumber \\
& \sim & 3\times 10^5\msun \left(\frac{v_{\rm line\, wind}}{4500\km\s^{-1}}\right)^2\left(\frac{\rph(t_{\rm spect})}{4\times 10^{14}\cm}\right) \, .
\end{eqnarray}
This inferred BH mass is remarkably similar to the BH mass we infer in \S\ref{sec:rt_MBH} via the independent argument of requiring that the disruption of a helium core take place outside the BH's innermost stable circular orbit.

We note that the half line-width (rather than the full line-width) is an appropriate measure of the wind velocity for the following reason.  In an optically thin outflow, we observe emission from gas moving towards us ($+v_{\rm line\, wind}$) and away from us ($-v_{\rm line\, wind}$), which together produce a broad emission line centered on the velocity of the central BH, whose width corresponds to twice the outflow velocity.  We find that this is the situation for the He II 4686 line of PS1-10jh, because the line is centered on the galaxy's lines \citep{gezari12}, and because the gas is marginally optically thin to He II 4686 photons (see Appendix \ref{app:heII_lines}).  If there were significant 4686$\ang$ absorption in addition to emission, the situation would be different: any observable absorption would take place along our line of sight, i.e., only in gas moving towards us; this would produce a P-Cygni line profile, which comprises blueshifted and redshifted emission superposed with blueshifted absorption (generally appearing simply as blueshifted absorption and redshifted emission). However, there was not enough He$^+$ gas in the $n=3$ energy level to produce significant 4686$\ang$ absorption because the gas was too cool (see Appendix \ref{app:heII_lines}); thus the He II 4686 appeared as a pure emission line.

We expect that if an early-time ultraviolet spectrum of PS1-10jh had been taken, the He L$\alpha$ line would have been observed. In contrast to the He II 4686 line, the He L$\alpha$ line {\it would} have exhibited a P-Cygni profile, because there was much more He$^+$ gas in the ground state $n=1$ than at $n=3$, and so the gas was highly optically thick to He L$\alpha$ absorption (see Appendix \ref{app:heII_lines}).  Absorption of He L$\alpha$ takes place between us and $\rph$, through gas moving towards us with velocities ranging from $\sim 0 - v_{\rm line\, wind}$; hence the absorption line would appear broad and blueshifted by $\sim 4500\km\s^{-1}$.  The gas outside $\rph$ all emits He L$\alpha$ as well, so a broad emission line (centered on the galaxy's lines) maybe have been visible superposed with the blueshifted absorption.  More speculatively, perhaps the width of the line (prior to $t_{\rm peak}$) decreased with time\footnote{Note that the observations of the He II 4686 line at $t_{\rm peak}+254\days$ were unable to measure its late-time width \citep{gezari12}. We speculate that the width of this line (and any other putative lines, such as He L$\alpha$), may have followed the velocity of the photosphere (perhaps decreasing prior to $t_{\rm peak}$ and increasing afterwards), but future detailed study of velocity evolution in line-driven TDE outflows will require more sophisticated models.} as $\Delta v \sim v_{\rm esc}(\rph)\propto \rph^{-1/2}\propto t^{-1/2}$.

Depending on the composition and ionization state of the gas, other ultraviolet resonance lines would have been visible as well, also with P-Cygni profiles (again, because they are due to absorption from the ground state $n=1$):  e.g., C III, Si IV, C IV.  (See \citealt{sq11} for detailed spectroscopic predictions of tidal disruption events in a slightly different context.)  Other Lyman series transitions of He$^{++}$ may also have been visible, and perhaps even a weak hydrogen Ly$\alpha$ line, if a trace amount of hydrogen was present. PS1-11af, a more recently discovered tidal disruption candidate, was observed with near-UV spectroscopy, and exhibited two broad and deep (unidentified) absorption lines at $\sim 2000-3000\ang$ 24 days after the event was discovered \citep{chornock14}; the physics was probably similar in this event.  Efforts are underway to obtain ultraviolet spectroscopy of disruption candidate ASASSN-14ae \citep{holoien} using the Hubble Space Telescope (S.B. Cenko et al., personal communication).

\section{Tidal disruption of a helium core}
\label{sec:hecore}

We have laid out evidence that PS1-10jh was too helium-rich to be the tidal disruption of a solar-type star.  As we explain further in \S\ref{sec:SNe}, the high fraction of energy output in radiation relative to kinetic energy (\S\ref{sec:energetics}), along with the behavior of the temperature and radius of the photosphere (\S\ref{sec:rad_photo}), are unlike the behavior of any known type of supernova.  Now we investigate the interpretation that PS1-10jh is the tidal disruption of the helium core of a red giant star \citep[proposed first by][]{gezari12}.

\subsection{Tidal disruption radius and black hole mass}
\label{sec:rt_MBH}

We calculated red giant models using the open-source stellar evolution code Modules for Experiments in Stellar Astrophysics \citep[version 4219 of MESA,][]{paxtonmesa}.  Our fiducial model has initial mass $1\msun$, initial metallicity $z=0.02$, and age $t_{\rm age} = 1.24\times 10^{10}\yr$.  
Red giant cores are very dense: the fiducial model has radius $\rcore \sim 0.03\rsun$ and mass $M_{\rm core} \sim 0.3\msun$, giving a mean density
$\bar{\rho}_{\rm core} \sim 2\times 10^4\g\cm^{-3}$.

Because the core's density is so high, the tidal force required to disrupt it is much greater than that for a solar-type star.
We estimate the core's tidal disruption radius (i.e., the distance from the BH at which the BH's tidal gravity can fully disrupt the core)\footnote{The tidal disruption radius of a star is $\rt = \xi R_{\ast} (\mbh/M_{\ast})^{1/3}$ where $\xi$ is typically order unity.  Simulations by \citet{guillochon13} of disrupting solar-type stars indicate that (depending somewhat on stellar structure) full disruption takes place for $\xi \approx 1$; we make the assumption that this result holds for red giant cores as well.  Note that \citet{macleod12}'s work on the disruption of red giant stars studies the disruption of red giants' low-density envelopes rather than the dense cores we are interested in here.},
\begin{eqnarray}
\rtcore & \sim & R_{\rm core} \left(\frac{\mbh}{M_{\rm core}}\right)^{1/3} \nonumber \\
& \sim & 2 \times 10^{11} \rcorerat \mcorerat^{-1/3} \mbhrat^{1/3} \cm \nonumber \\
& \sim & 3 \rs \rcorerat \mcorerat^{-1/3} \mbhrat^{-2/3} \, , \label{eq:rtcore}
\end{eqnarray}
where $\rs = 2G\mbh/c^2$ is the BH's Schwarzschild radius.

The requirement that the core be fully disrupted to produce an observable event places significant constraints on the mass of the BH.
If the BH were too massive\footnote{This upper mass limit would presumably be somewhat larger if we account for spinning BHs; see \citealt{kesden}. Work by \citet{bogdanovic_cheng} also suggests a weaker upper limit to $\mbh$ by allowing tidal heating of the core to reduce the core's density; see \S\ref{sec:futurework}.} $\mbh \gtrsim 10^6\msun$, the core's tidal radius would lie inside the BH's event horizon ($\rtcore \lesssim \rs$) and the BH would swallow the core whole rather than tidally disrupting it.  If the star's orbital pericenter lies outside the event horizon but inside the BH's ISCO ($r_{\rm ISCO} = 3\rs$ for a non-spinning BH), much of the gas may be accreted by the BH shortly after disruption (see \citealt{haas} for simulations of white dwarf disruption at different pericenter distances). It is likely that this scenario would not produce much emission either, although future work incorporating general relativistic effects, accretion physics, and radiation will be necessary to make firmer predictions.
Based on these arguments, we infer that $\mbh \lesssim 10^6\msun$---and more likely $\mbh \sim 2\times 10^5\msun$ or less.  We adopt this value of $\mbh = 2\times 10^5\msun$ as our fiducial BH mass, with a fiducial tidal disruption radius of $\rtcore \sim 2\times 10^{11}\cm$.

The inferred bolometric luminosity $L_{\rm bol} \gtrsim 2\times 10^{44}\erg\s^{-1}$ \citep{gezari12} is thus about 4 times the Eddington luminosity appropriate for doubly-ionized helium gas,
\begin{eqnarray}
L_{\rm Edd,He} & = & \frac{4\pi G\mbh c}{\kappa_{\rm es,He}} \\
& = & 5\times 10^{43}\mbhrat \erg\s^{-1} \, . \nonumber
\end{eqnarray}
%
This indicates that radiation pressure plays a significant role in the dynamics of PS1-10jh, as we shall see further shortly.

\subsection{Fallback}\label{sec:hecore:fallback}
\begin{figure}
\centerline{\epsfig{file=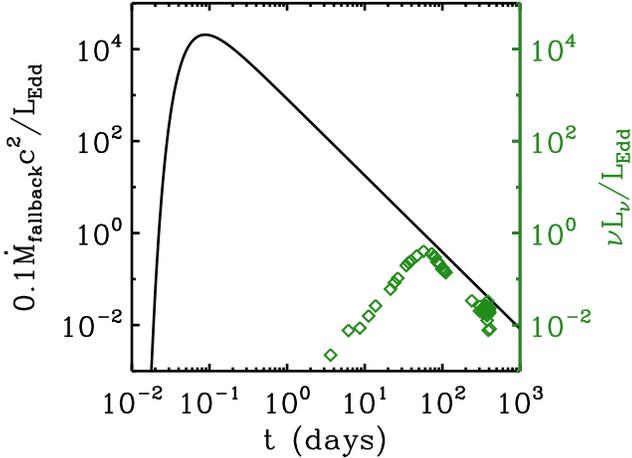, width=3.5in}}
\caption{Black solid curve:  Rate of energy generation in gas falling back to pericenter in units of the Eddington luminosity, $0.1\mdotfb c^2/L_{\rm Edd}$, for our fiducial model of a red giant core ($M_{\rm core}=0.3\msun$ and $\rcore = 0.03\rsun$), assuming $\mbh = 2\times 10^5\msun$ and $\rtcore = 2\times 10^{11}\cm$.   The fallback rate is hugely super-Eddington shortly after disruption, and remains super-Eddington for months.  Green diamonds: $g$-band data for PS1-10jh \citep{gezari12} in units of the Eddington luminosity.   That the light curve peaks $\sim 3$ orders of magnitude later than the fallback rate peaks indicates that the photons generated in the fallback shock get trapped (though more photons may be generated in a shock at larger radius; see \S\ref{sec:adiabatic_cooling}). Photons cannot escape until much later when the flow becomes optically thin (at $t\sim 67\days$). After that time, the light curve approximately follows the energy generation rate. \label{fig:mdotomdotedd}}
\end{figure}

In the now ``classical'' picture \citep[e.g.,][]{lacy,rees}, as the stellar core disrupts, half of its gas becomes bound to the BH.  The bound gas departs the BH on highly eccentric elliptical orbits. The most bound gas\footnote{We assume for simplicity that the star is maximally spun up during disruption.} has energy
$\mathcal{E} \sim (3 G\mbh/\rtcore)(\rcore/\rtcore)$ \citep[e.g.,][]{lacy,li02}, so that its apocenter is
\begin{eqnarray}\label{eq:Rapo}
\rapo & \sim & \frac{\rtcore^2}{3\rcore} \\
& \sim & 6 \times 10^{12} \left(\frac{\rtcore}{2\times 10^{11}\cm}\right)^2
\left(\frac{\rcore}{0.03\rsun}\right)^{-1} \cm \nonumber
\end{eqnarray}
(roughly twice the semi-major axis because the orbit is highly eccentric).
This most tightly bound gas begins returning to pericenter after the ``fallback time''
\begin{eqnarray}
\tfall & \sim & \frac{2\pi}{6^{3/2}}\left(\frac{\rtcore}{\rcore}\right)^{3/2}\left(\frac{\rtcore^3}{G\mbh}\right)^{1/2} \nonumber \\
& \sim & 7\times 10^3 \mbhrat^{-1/2}\rtcorerat^{3} \nonumber \\
& & \times \rcorerat^{-3/2} \s \, ,
\label{eq:tfall}
\end{eqnarray}
%
%
\citep[e.g.,][]{rees, li02}---only about 2 hours.
Bound gas continues ``falling back'' to the BH at a rate
\begin{equation}
\mdotfb \approx \frac{1}{3}\frac{\mcore}{\tfall}\left(\frac{t}{\tfall}\right)^{-5/3}
\label{eq:mdotfallback}
\end{equation}
\citep{rees, phinney89}\footnote{There may be slight or moderate deviations from this canonical form depending on details of the core's density profile: see \citet{lodato09} and \citet{guillochon13} for results for solar-type stars.}, depicted by the solid black curve\footnote{\label{ft:Gamma} In Figure \ref{fig:mdotomdotedd}, we plot the fallback rate from equation (\ref{eq:mdotfallback}) multiplied by a factor $3/\Gamma(1/3) \times \exp[ -(\tfall/t)^2 ]$ to approximate the onset of fallback; this shape is similar that of \citet{evkoch}'s numerical result for the disruption of a solar-type star.  This diminishes the maximum fallback rate by a factor of several.} in Figure \ref{fig:mdotomdotedd}.
Note that most of the mass falls back at early times:  within a time $t\sim 10 \tfall \sim 1$ day after disruption, 80\% of the bound gas has fallen back to pericenter; already half of it has fallen back after only $t\sim 3\tfall$.
We argue in \S\ref{sec:simplemodel} that this could be an explanation for the shell-like geometry of the expanding gas.

When returning gas reaches the vicinity of the BH, it is expected to shock on itself \citep[e.g.,][]{rees,evkoch,kochanek94,guillochon09} and convert orbital kinetic energy into radiation energy with some efficiency $\eta$.  If this radiation were able to escape immediately, the bolometric light curve of the event would be proportional to the mass fallback rate $\mdotfb$.
The peak luminosity of the event would be
\begin{eqnarray}
L(\tfall) & \sim & \eta\dot{E}_{\rm fallback} \sim 0.1\mdotfb(\tfall) c^2 \\
& \sim &  10^{48} \mbhrat^{1/2}\mcorerat \nonumber \\
& & \times \rtcorerat^{-3}\rcorerat^{3/2} \erg\s^{-1} \, , \nonumber
\end{eqnarray}
  with an effective temperature of 
\begin{eqnarray}
kT_{\rm eff} & \sim & k\left(\frac{0.1\mdotfb(\tfall)c^2}{4\pi\rtcore^2\sigma_{\rm SB}}\right)^{1/4} \nonumber \\
& \sim & 1 \mbhrat^{1/8}\mcorerat^{1/4}\rtcorerat^{-5/4} \nonumber \\
& & \times \rcorerat^{3/8} \keV 
\end{eqnarray}
%
(assuming the radiation had time to thermalize), lasting for a timescale of $\sim {\rm few}\times \tfall\sim$ hours.  Here $\sigma_{\rm SB}$ is the Stefan-Boltzmann constant, and $\dot{E}_{\rm fallback}$ is the rate at which orbital kinetic energy is dissipated in the shock at pericenter at time $\tfall$, assumed to be $\sim 0.1\mdotfb(\tfall)c^2$ for some reasonable value of the efficiency factor $\eta$.
This timescale, luminosity, and temperature are dramatically different from those observed for PS1-10jh.

Recent numerical work \citep[e.g.,][]{shiokawa15,bonnerot15} suggests modifications to this ``classical'' picture: the shock at pericenter may be rather weak compared to stream-crossing shocks that occur near apocenter ($\rapo$: see eq. \ref{eq:Rapo}). The effect is to prolong the duration of peak accretion, by a factor of 3-10, and to reduce the peak accretion rate. However, even with the prolonged timescale, the accretion timescale is still two orders of magnitude shorter than the observed duration of peak luminosity.  Additionally, the shock at pericenter may be strong nevertheless for disruptions close to the ISCO, as PS1-10jh seems likely to be. We continue here assuming fallback times and accretion rates that correspond to a shock at pericenter rather than apocenter, but discuss these issues further in \S\ref{sec:extra_shock}.

\subsection{Trapped radiation}\label{sec:hecore_trap}
In Figure \ref{fig:mdotomdotedd}, we plot the $g$-band light curve observed for PS1-10jh (green diamonds) over the classical estimate for the energy generation rate at the shock, $\sim 0.1 \mdotfb c^2$ (see Footnote \ref{ft:Gamma}).  The different shapes are striking:  the light curve peaks $\sim 3$ orders of magnitude later in time than the energy generation rate, and the bolometric light curve's peak brightness is $\sim 3$ orders of magnitude below the peak energy generation rate (assuming that the $g$-band luminosity is $\sim 10\%$ of the bolometric luminosity).  Interestingly, the light curve's decay does approximately follow the energy generation rate.

We thus propose that, instead of immediately escaping, most of the radiation produced either in the shock at pericenter or by the accretion disk was {\it trapped} at small radii by electron scattering.  The optical depth in the flow at pericenter would initially have been enormous,
\begin{eqnarray}\label{eq:taufall}
\tau(3\tfall) & \sim & \frac{\kappa_{\rm es,He}(\mcore/4)}{4\pi\rtcore^2} \\
& \sim & 6 \times 10^7 \mcorerat \rtcorerat^{-2} \gg 1 \, ,  \nonumber
\end{eqnarray}
%
(where we are approximating the flow as spherical).
If the stellar gas around the BH remained static, the time for photons to diffuse out would be
\begin{eqnarray}\label{eq:tdiff}
t_{\rm diff}(\rtcore) & \sim & \frac{\tau(3\tfall)\rtcore}{c} \\
& \sim & 10 \mcorerat \rtcorerat^{-1} \yr \, , \nonumber
\end{eqnarray}
%
much longer than the fallback time or dynamical time (or rise time of the observed light curve). 

We propose that the trapped radiation pushes gas outward: the radiation produced via fallback
is easily energetic enough to unbind a fraction of the originally bound half of the stellar gas, since the fallback rate shortly after fallback begins (say, at time $3\tfall$) is orders of magnitude above the Eddington rate (see Figure \ref{fig:mdotomdotedd}),
\begin{eqnarray}\label{eq:mdotomdotedd}
\frac{\mdotfb}{\dot{M}_{\rm Edd}} & \sim & 8\times 10^3 \mbhrat^{-1/2} \mcorerat  \nonumber \\
& & \times \rtcorerat^{-3}\rcorerat^{3/2} \, .
\end{eqnarray}
%
The observations directly show that there is enough energy available to lift a substantial amount of mass out of the potential well of the black hole; the energy we observe in radiation is sufficient to lift $\sim 0.01\msun$ from a radius of several $\rs$ (c.f. eqs. \ref{eq:Ep} and \ref{eq:Eradbolout}).

\section{Interpretation}
\label{sec:interpretation}

We now draw together the arguments from the previous sections to propose the following picture for what took place in PS1-10jh.

\subsection{Simple model}
\label{sec:simplemodel}

The helium core of a red giant star was disrupted by a relatively low-mass BH ($\mbh \lesssim 2\times 10^5\msun$).  In the hours following disruption, newly-bound stellar gas completed elliptical orbits around the BH and fell back to pericenter at a highly super-Eddington rate (eq. \ref{eq:mdotomdotedd}), shocking on itself and quickly generating huge amounts of radiation.  The gas was incredibly optically thick (eq. \ref{eq:taufall}), and so essentially all of the radiation was trapped in the gas by electron scattering.

The radiation contained enough energy to gravitationally unbind a significant fraction of the stellar gas that had fallen back to the BH:  about $10\%$ ($0.01\msun \sim 0.1(M_{\rm core}/2)$) was driven out
in an optically thick shell at $v\sim 1000\km\s^{-1}$, whose expanding edge was observed as the expanding photosphere $\rph$.  While the shell remained optically thick, photons were advected out with the gas faster than they could make progress diffusing out.  Radiation was mostly only able to escape (and reach our telescopes) once the shell had expanded enough that the density had fallen enough to allow the shell to transition from optically thick to optically thin.  This transition took place around $t\sim 67\days$ after disruption, producing the observed peak of the light curve (long after the peak of accretion).

Outside the photosphere for Thomson scattering, a fraction of the radiation was trapped by resonance lines (e.g., helium Lyman $\alpha$), which accelerated a tiny wisp more gas to velocities $\sim 4500\km\s^{-1}$.  This outermost gas was photoionized by the radiation coming from the Thomson photosphere, and produced the broad He II recombination lines that were observed (which serve as clues to the composition of the disrupted object).

Eventually the expanding shell became optically thin, and from then on we could see through to gas inside it.  Although gas particles continued to expand outwards, the location of the photosphere now moved {\it inwards} as the density of the inner gas fell and allowed us to see deeper and deeper inside: see \S\ref{sec:geometry} and Figure \ref{fig:rhotaushell}.  We hypothesize that the shell geometry was present because of the very short timescale ($\Delta t \sim {\rm few}\times\tfall \ll t_{\rm peak}$) over which most of the mass fell back to pericenter and most of the radiation was generated.  The gas interior to the shell fell back to pericenter at later times $t\gtrsim {\rm few}\times\tfall$: it was subsequently driven out by radiation pressure, and we speculate that it may have had more of a wind geometry $\rho \propto r^{-2}$, as proposed by \citet{sq09}.  

Now that the total column from the BH out through the flow was so much smaller, most of the radiation generated close to the BH could escape, rather than being swallowed. The radiation continued to be reprocessed by the outer gas (hence its cool $3\times 10^4\K$ temperature), but not nearly as heavily as while the shell was optically thick. Thus we see that the photometric measurements at these late times are similar to $0.1 \mdotfb c^2$ (Figure \ref{fig:mdotomdotedd})\footnote{We should more properly compare the bolometric luminosity $L_{\rm bol}$ with $0.1 \mdotfb c^2$, but as the temperature remains roughly constant, the $g$-band emission remains a constant fraction $\sim 0.1$ of $L_{\rm bol}$. }.

\subsection{Energy comparisons}
\label{sec:interp_energy}

This simple model explains some of the energy scales we inferred in \S\ref{sec:energetics}.

\begin{enumerate}
\item As the shell of stellar gas expanded, trapped radiation energy was converted to kinetic and potential energy of the gas particles.
Conserving energy, we'd expect to find that the initial radiation content of the shell (before it expanded) $E_{\rm rad,0}$ is equal to the total energy in the gas when the remaining radiation was finally able to escape,
\begin{equation}
E_{\rm rad,0} \sim E_{\rm rad,bol,out}+E_{\rm K}+\Delta E_{\rm P} \, ,
\end{equation}
Our observation that $E_{\rm rad,bol,out}\sim E_{\rm K}+\Delta E_{\rm P}$ suggests that the initial radiation content of the shell was a few times the quantity of radiation that escaped ($E_{\rm rad,0}\sim {\rm few}\times E_{\rm rad,bol,out}$).  We discuss this further in \S\ref{sec:adiabatic_cooling}.

\item The total energy released by the disruption $E_{\rm dis}$ was almost an order of magnitude greater than the total observed energy of the shell (eq. \ref{eq:Etot}),
\begin{equation}
E_{\rm tot,out} = \left(E_{\rm rad,bol,out}+E_{\rm K}+\Delta E_{\rm P}\right) \sim 0.2 E_{\rm dis} \, ,
\end{equation}
assuming that $\eta \sim 10\%$ of the accretion energy was converted to radiation (see eq. \ref{eq:Edis}):
Evidently, the remaining $\sim 80\%$ of the radiation energy liberated by the disruption was quickly swallowed by the BH, presumably entrained in the $\sim 90\%$ of the gas that was not blown away in the shell. (Note that we infer that the fraction of gas that {\it was} blown away in the shell is $\sim 0.01\msun/ 0.15\msun \sim 7\%$, discussed further in \S\ref{sec:discussion}.)

Simulations of accretion physics often measure a quantity we'll call $\eta_{\rm esc}$, the efficiency with which mass fed to the BH is converted to radiation that then manages to escape from the flow to infinity.  This quantity differs from what we've called $\eta$, the efficiency with which mass fed to the BH is converted to {\it any} radiation---much of that radiation can then be swallowed by the BH and not contribute to $\eta_{\rm esc}$.
Our estimate for $\eta_{\rm esc}$ in PS1-10jh, from comparing the observed quantity $E_{\rm tot,out}$ to (half) the mass of the stellar core (i.e., $E_{\rm dis}/\eta$ in eq. \ref{eq:Edis}), is
\begin{equation}\label{eq:eta}
\eta_{\rm esc} \sim 0.02 \left(\frac{E_{\rm tot,out}}{5\times 10^{51}\erg}\right) \left(\frac{M_{\rm core}}{0.3\msun}\right)^{-1} \, .
\end{equation} 
This result is intriguingly similar to results of the recent 3-dimensional radiation magnetohydrodynamic simulations by \citet{jiang14} who find $\eta_{\rm esc} = 0.045$, discussed further in \S\ref{sec:discussion}.

\item In supernovae, the majority of the radiation energy is typically converted to kinetic energy, leading to the observed $E_{\rm rad,out} \sim 10^{49}{\rm erg}\ll E_{\rm K}\sim 10^{51}{\rm erg}$ \citep{2009ARA&A..47...63S}.  PS1-10jh contrasts starkly with this: $E_{\rm K}\sim 10^{47}{\rm erg} \ll E_{\rm rad,out}\sim 10^{51}{\rm erg}$ (\S\ref{sec:energetics}).  We suggest that the kinetic energy was much smaller than the radiation energy for two reasons. First, in contrast to supernovae, much of the energy initially stored in the radiation was converted to {\it potential} energy instead of kinetic energy---photons had to do work against the gravitational potential of a massive BH, work the photons don't have to do in a supernova.  Another reason for $E_{\rm K} \ll E_{\rm rad,out}$ could be that the shell experienced a shock at large radius, converting a substantial fraction of the kinetic energy into photons; see \S\ref{sec:adiabatic_cooling}.  This difference in energetics provides strong evidence that PS1-10jh is not a type of supernova. We offer a fuller comparison with supernovae later in \S\ref{sec:SNe}. 

\item It seems significant that the outflowing shell of gas was only {\it barely} unbound from the BH, as we deduce in two ways: (1) The kinetic energy is tiny compared with the potential energy ($E_{\rm K} \ll \Delta E_{\rm P}$: \S\ref{sec:energetics}), and (2) The outflow velocity is similar to the escape velocity at large distance from our fiducial BH (eq. \ref{eq:vesc}).  By contrast, winds from AGN or O-stars typically have velocities several times the escape velocity from the radius where they are launched \citep[e.g.,][]{murray95,lamers99}; if PS1-10jh had followed suit, its expansion velocity would have been $\sim 30$ times greater.  Did some process regulate radiation pressure to cancel gravity almost exactly?

\item The model does not explain the observed constant $T_{\rm eff}$ of the photosphere.  The large size of the photosphere relative to the tidal radius $\rtcore$ suggests that the radiation cools by scattering and/or advection through optically thick gas between the location where the radiation is generated (the shock or inner regions of the accretion disk) and the location where the radiation begins to free-stream.  But as we point out in equation (\ref{eq:adiabatic}), adiabatic expansion would suggest that the photosphere cooled as the gas expanded, which was not observed.  In \S\ref{sec:adiabatic_cooling} and \S\ref{sec:discussion}, we discuss further ideas about what sets the temperature.

\end{enumerate}

\subsubsection{Adiabatic cooling? -- An additional shock}
\label{sec:adiabatic_cooling}
The final energy comparison of this section addresses the question of adiabatic cooling of the radiation.
Most of the radiation escaped close to the time $t_{\rm peak}$, when the shell became optically thin, so $E_{\rm rad, out}$ is approximately the radiation content of the flow at that time:  $E_{\rm rad}(t_{\rm peak}) \sim E_{\rm rad,out}$.
We have argued that at times earlier than this, most of the radiation was trapped by Thomson scattering within the expanding flow.  As the flow expands, we expect the energy in radiation to decrease due to adiabatic cooling \citep[e.g.,][]{sq09, rossi09}.  The temperature of an adiabatically expanding photon gas scales as $T \propto V^{-1/3} \propto \redge^{-1}$, where $V$ and $\redge$ are the volume and radius of the flow; the energy in radiation therefore scales as
\begin{equation}\label{eq:adiabatic}
E_{\rm rad} \sim a T^4 V \propto \redge^{-1} \, .
\end{equation}
We can use this relation to work our way backwards in time and infer the radiation content of the gas early on. Supposing the flow originated close to the tidal radius for a disrupting red giant core $\sim 2\rtcore$ (see \S\ref{sec:rt_MBH}), we'd estimate that the radiation energy was initially\footnote{Even using equation (\ref{eq:adiabatic}) to estimate the radiation energy at the time of first detection $t_1$ gives an uncomfortably high $E_{\rm rad}(t_1) \sim 2 \times 10^{52} \erg$.} a whopping
\begin{equation}
E_{\rm rad,0} \sim 3 \times 10^{54} \left(\frac{\rph(t_{\rm peak})}{6\times 10^{14}\cm}\right) \left(\frac{\rtcore}{2\times 10^{11}\cm}\right)^{-1}  \erg \, ,
\end{equation}
fully 100 times larger than the energy released in the disruption of a red giant core (eq. \ref{eq:Edis}).  Where could all that energy have come from?
 
Disrupting a much more massive star could solve the problem, as $E_{\rm dis}$ would be correspondingly larger; however, postulating the disruption of an exceedingly rare $\sim 30\msun$ core seems far-fetched. 
Instead, we argue that the expansion was {\it not} adiabatic.

For comparison, in a typical supernova explosion, expansion {\it is} adiabatic and homologous: concentric shells expand at different velocities such that they do not interact with each other, and most of the radiation content of the gas is converted to kinetic energy (so that $E_{\rm rad}\ll E_{\rm K}$) as the flow expands in an orderly fashion \citep{arnett82}. The class of superluminous supernovae, however, which have $E_{\rm rad}\sim 10^{51}{\rm erg}$,  appear to require non-adiabatic behavior \citep{quimby}.

In contrast to the standard supernova picture, we suggest that the flow in PS1-10jh underwent a shock or shocks at large radius from the BH, similar to a mechanism proposed for making superluminous supernovae so bright \citep[e.g.,][]{gal-yam12}; see \S\ref{sec:SNe}.  In superluminous supernovae, it is believed that the high-velocity ($\sim 10,000\km\s^{-1}$) ejecta encounter a much slower wind expelled by the star earlier. In PS1-10jh, perhaps a weak shock at apocenter \citep{shiokawa15, piran15} provided gas for the shell to shock on (see \S\ref{sec:extra_shock}).  However, the slow speed of PS1-10jh's expanding shell ($\sim 1000\km\s^{-1}$) would likely have hindered the effectiveness of such an external shock. We speculate instead that there could have been internal shocks (similar to models for gamma-ray bursts; e.g., \citealt{grb_internal}): deep inside the shell, perhaps more gas was accelerated outward from small radii (perhaps by radiation pressure), at speeds that (for some reason) were faster than the shell's expansion. As this faster inner gas caught up to and collided with the slower-moving shell, the resulting shocks would convert the wind's kinetic energy to radiation energy, producing the relatively large quantity of radiation observed ($E_{\rm K} \ll E_{\rm rad}$). It would also impart momentum to the shell, which could help explain why the observed expansion velocity was constant, despite the fact that the central black hole's gravity should have continuously decelerated the gas (see Figure \ref{fig:kinematics}).
This continual reheating of the gas could perhaps have maintained the temperature of the photosphere as well.  Investigating these ideas further likely will require radiation hydrodynamical simulations.

\section{Comparison with supernovae}
\label{sec:SNe}

\citet{gezari12} argue against a supernova interpretation for PS1-10jh because of its long-lasting UV emission:  GALEX still observed UV emission fully 375 (rest-frame) days after the light curve's peak.  They argue that a core-collapse supernova would have cooled to $\sim 6000\K$ a month after the explosion; the lack of recent star formation in the host galaxy also makes a core-collapse supernova unlikely.  Here we offer a further comparison between the properties of PS1-10jh and those of supernova (SN) explosions.

\subsection{Photosphere and energy}
In \S\ref{sec:rad_photo}, we calculated the radius of the photosphere of PS1-10jh by comparing the temperature with the blackbody luminosity (eq. \ref{eq:R(t)}); we found that $\rph$ increased with time for $\sim 67\days$, reached a maximum of $\rph \sim 6\times 10^{14}\cm$, and then notably {\it decreased} with time (Figure \ref{fig:roft}).  Because the effective temperature remained roughly constant, the expanding and receding photosphere caused the light curve to rise and fall. This scenario differs strongly from supernovae, where the radius of the photosphere generally increases with time up through the peak of the light curve and beyond; supernova photospheres only begin to recede once the whole flow has become optically thin \citep[e.g.,][and references therein]{arnett82, chornock14}.  For supernovae, the fading light curve is not due to a receding photosphere, but rather to the temperature of the gas falling (mostly because of adiabatic expansion of the gas).  \citet{chornock14} calculate $\rph$ for PS1-11af and find that its photosphere rises and falls as well; they also point out that such behavior clearly distinguishes that event from supernovae.

This difference presumably comes about because the outflowing gas in PS1-10jh is less massive ($\sim 0.01\msun$; eq. \ref{eq:mshell}) than the ejecta in a typical supernova explosion ($\gtrsim 1\msun$; \citealt{2009ARA&A..47...63S}): the column density in the expanding shell falls enough to become optically thin much sooner in PS1-10jh than in a SN.  Another difference is that in a supernova, the photosphere begins to recede at late times when it has cooled enough that the gas recombines and Thomson scattering ceases to provide opacity.  PS1-10jh is {\it not} seen to cool, and so Thomson scattering remains important.

We note also that while the peak luminosity and total energy emitted from superluminous supernovae (SLSNe: see next section) are similar to PS1-10jh, the photosphere for PS1-10jh is a factor of $\sim 10$ smaller than that for SLSNe (e.g., PTF09cnd has $\rph \sim 5\times 10^{15}\cm$ at peak luminosity; \citealt{quimby}).  This observational difference is because PS1-10jh is significantly hotter than SLSNe, and so needs a smaller photosphere to produce the same luminosity and energy output.

Also at odds with supernovae is our finding that the kinetic energy in PS1-10jh is much smaller than the radiation energy ($E_{\rm K} \ll E_{\rm rad}$; \S\ref{sec:energetics}); supernovae typically exhibit $E_{\rm rad} \ll E_{\rm K}$ \citep[e.g.,][]{arnett82,2009ARA&A..47...63S}.  We suggested in \S\ref{sec:interp_energy} that this difference arises because the radiation in PS1-10jh was mostly converted to potential energy instead of kinetic energy, since photons had to do work against the gravitational potential of a massive BH (which they don't have to do in a supernova).  We also suggested in \S\ref{sec:adiabatic_cooling} that PS1-10jh may not expand adiabatically and homologously as SNe typically do \citep{arnett82}; kinetic energy may have been converted back to radiation energy through shocks in the expanding gas.  This would also have kept PS1-10jh from cooling as SNe do.

\subsection{Light curves, spectra, and super-luminous supernovae}

\begin{figure}
\centerline{\epsfig{file=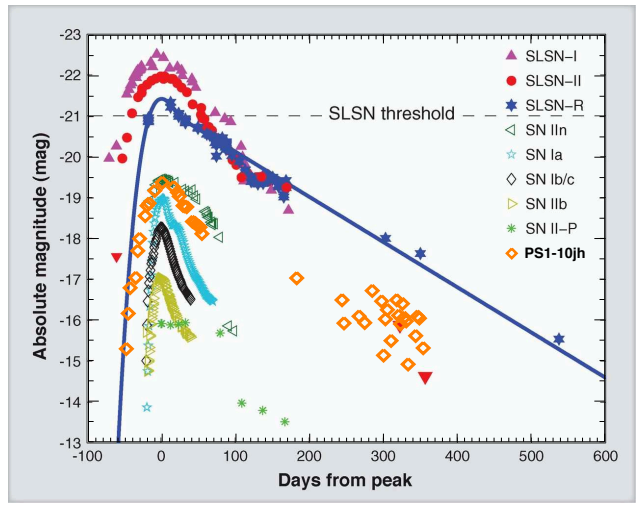, width=3.5in}}
\caption{Figure 1 from \citet{gal-yam12}, with data for PS1-10jh overplotted (orange diamonds). All supernova data is in the observed $R$ band \citep[see][]{gal-yam12}; data for PS-10jh is in the observed $r$ band \citep[from][]{gezari12}.  The dashed line shows the threshold for classification as a super-luminous supernova (SLSN).  PS1-10jh appears fainter than SLSNe because it is hotter, so less of its radiation emerges in the $r$-band, despite a peak bolometric luminosity similar to SLSNe.  PS1-10jh fades more slowly than most SNe, looking most similar in light-curve shape to SLSN-R events (but these have completely different spectral properties; see text).   \label{fig:GalYamcomp}}
\end{figure}

We compare the light curve of PS1-10jh with those of supernovae in Figure \ref{fig:GalYamcomp}.  We focus particularly on a growing class of supernovae called ``super-luminous supernovae'' (SLSNe):  similar to PS1-10jh, these events reach peak luminosities $\sim 10^{44}\erg\s^{-1}$ and radiate total energies $\sim 10^{51}-10^{52}\erg$, 10 or more times brighter than ``common'' SNe \citep{gal-yam12}.  Data for PS1-10jh is plotted in orange diamonds over \citet{gal-yam12}'s Figure 1.

Figure \ref{fig:GalYamcomp} and comparison with spectra in \citet{gal-yam12} show that PS1-10jh differs substantially from SNe, including SLSNe.  Despite its large bolometric luminosity that would classify it as a SLSN, PS1-10jh falls below the threshold when only $r$-band light is considered, because PS1-10jh is significantly hotter and most of its radiation is emitted at ultraviolet wavelengths; UV emission in SNe is typically suppressed by line blanketing as well. PS1-10jh fades more slowly than most supernovae, looking most similar in decline to \citet{gal-yam12}'s SLSN-R class, which is believed to be powered by radioactive decay of nickel-56; however, the optical spectra of SLSN-R events show many metal absorption lines, completely absent in the spectra of PS1-10jh.  SLSN-II events fade more quickly than PS1-10jh, and have spectra that contain strong hydrogen emission lines (unlike PS1-10jh); these events are thought to be explosions that take place inside thick hydrogen envelopes (either inside previous expulsions from a dying star, or inside an incredibly bloated star itself) which are heated and ultimately allow the radiation to diffuse out.  SLSN-I events, driven perhaps by a combination of radioactive decay and interaction with a thick (hydrogen-free) envelope, also fade more quickly than PS1-10jh.  They are typically very blue and luminous in the ultraviolet (like PS1-10jh), but their UV emission lasts a few months, not a year (as PS1-10jh does).  Their spectra show no hydrogen, but they also show no prominent helium close to the light curve's peak, and do show broad absorption lines due to metals, which PS1-10jh lacks.

In this framework, the best bet for a supernova interpretation would probably be an event similar to a SLSN-II, in which an explosion took place inside a thick shell of gas.  However, the shell would have to be helium-rich rather than hydrogen-rich---perhaps a star that shed its hydrogen envelope longer in the past, and shed its helium envelope shortly before explosion. This is similar to, e.g., the scenario proposed for a set of SLSNe in \citet{quimby}; however, in their case, the spectra showed neither hydrogen nor helium (and instead showed other elements\footnote{In the optical, \citet{quimby}'s SLSNe show strong O II absorption lines; PS1-10jh was too hot for O$^{++}$ to be present in significant quantities, so its spectra show no O II.}).  Moreover, the requisite shell proposed by \citet{quimby} weighs many solar masses and is produced by the violent pulsations of an extremely massive dying star ($\mstar \sim 100\msun$), highly unlike the puny $0.01\msun$ shell we infer for PS1-10jh.

Velocities inferred for SLSNe's expanding shells and shockwaves through the shells don't obviously match the properties of PS1-10jh either:  e.g., \citet{quimby}'s events (type SLSN-I in \citealt{gal-yam12}'s nomenclature) have photosphere expansion speeds of $\sim 10^4\km\s^{-1}$ rather than $10^3\km\s^{-1}$, and SLSN-II events show H$\alpha$ linewidths of $\sim 4000\km\s^{-1}$ \citep[e.g.,][]{smith08,gal-yam12}, somewhat slower than $\sim 9000\km\s^{-1}$ observed for PS1-10jh.  We note that the ratio of velocities of PS1-10jh's He II 4686 line (likely the velocity of a line-driven wind) to its photospheric expansion is $\sim 5$, while this ratio for SLSNe is $\sim 0.3$.  This may be because SLSNe quickly become too cool to continue driving a wind.

For a final comparison, even the spectrum of an ``unusual helium-rich'' supernova, believed to have lost its hydrogen envelope prior to explosion \citep{kawabata}, looks nothing like PS1-10jh: the supernova shows He I lines rather than He II, and they are in absorption rather than in emission.

Thus we conclude that PS1-10jh differs significantly in multiple important ways from known types of supernovae.

\section{Discussion}
\label{sec:discussion}

\subsection{Summary}
\label{sec:summary}

PS1-10jh is a transient discovered by the Pan-STARRS Medium Deep Survey in 2010 and identified as a candidate tidal disruption event.  This event has one of the best-sampled light curves of a TDE to date, including the rise to peak luminosity followed over two months.  Optical and UV photometry show that the temperature of the photosphere was $\sim 3\times 10^4\K$ and remained fairly constant over time.  Optical spectroscopy showed two He II recombination lines in emission, but no hydrogen (or any other) lines.
We have investigated the nature of PS1-10jh, making arguments based on both the continuum and spectroscopic observations.  Here we summarize and discuss our conclusions, and point out directions for future work.

\begin{enumerate}
\item The observed constant temperature $\teff \sim 3 \times 10^4\K$ and the observed light curve, when combined, yield the radius of the photosphere $\rph$ as a function of time, assuming that the emission has a blackbody shape.  The photosphere is seen to grow, reach a maximum of $\sim 6\times 10^{14}\cm$, then decline.  The growth is close to linear in time, increasing with a velocity $v_0 \sim 1000\km\s^{-1}$.
\item We interpret the growth of the photosphere to indicate the expanding edge of a shell of gas. The photosphere reaches its maximum size when the shell has an optical depth of unity; thereafter, the shell is optically thin and we see through it to gas inside.  This constraint allows us to estimate the mass of gas in the shell, $\sim 0.01\msun$.
\item The mass and velocity estimates provide an estimate for the kinetic energy of the gas, $E_{\rm K} \sim 10^{47}\erg\s^{-1}$.  \citet{gezari12} estimated the total energy emitted in radiation $E_{\rm rad,bol,out}\sim 2\times 10^{51}\erg$.  We showed that the luminosity in the observed bands (with no assumption about the shape of the spectral energy distribution, i.e., without assuming the emission is thermal) was $\gtrsim 6\times10^{50}\erg$. The ratio of kinetic to radiation energy is thus $\sim 10^{-4}$.  This is a factor of a million smaller than the typical ratio for supernovae, $\sim 10^2$ \citet{2009ARA&A..47...63S}. 
\item He II emission lines are produced by recombination of He$^{++}$ ions with electrons. Photoionization calculations demonstrate that the observed lower limit of the ratio of He II $4686\ang$ to H$\alpha$ implies that the emitting gas is helium-rich; the gas does {\it not} have solar composition.  
\item Based on these photoionization results, a likely possibility is that the disrupted object was a red giant core whose envelope was previously stripped.  Such an object is dense enough that it requires a small BH ($\mbh \lesssim 2 \times 10^5\msun$) to disrupt the object outside the BH's innermost circular orbit.
\item The (half) linewidth of the He II line and contemporaneous measurement of the radius of the photosphere together imply the mass of the central object, $\sim 3\times 10^5\msun$.  This estimate is close to the BH mass we inferred by independent argument in the previous point. 
\item From the luminosity of the He II 4686 line, we infer that the emission measure of the gas producing the line is $(EM) \sim 3 \times 10^{65}\cm^{-3}$, similar to the independent estimate of the emission measure we made based on the radius of the photosphere (from continuum observations).
\item We infer that gas from the disrupted star fell back to the BH at a rate much larger than the Eddington rate for an extended period (hours to a day) following disruption.  Photons were trapped in the highly optically thick gas, and radiation pressure drove the expansion of the gas shell; the shell was gravitationally unbound from the BH, but (surprisingly) only barely.  Trapping of radiation ensures that the light curve of the event is far from simply proportional to the mass fallback rate (Figure \ref{fig:mdotomdotedd}): future investigations of tidal disruption {\it must} account for radiation pressure.
\item Outside the photosphere to Thomson scattering, resonance lines (such as He L$\alpha$) continued to trap a fraction of radiation and drive a further wind, slightly faster than the expansion of the shell.
We encourage observers to seek ultraviolet spectroscopy of future tidal disruption candidates, and predict that their ultraviolet spectra will show broad resonance lines, blue-shifted and seen in absorption (or perhaps having a P-Cygni profile), as would characterize an outward-moving wind \citep[also predicted in][]{sq11}. We also encourage contemporaneous X-ray observations, to better constrain the incident SED for modeling photoionization in the line-emitting gas.
\item The emitted radiation $E_{\rm rad,bol,out}\sim 2\times 10^{51}\erg$ is only 10 times less than the energy released in the disruption of a helium core, $\sim 3 \times 10^{52}\erg$.  This implies that the radiation we observe did {\it not} start out at a small radius $\sim \rtcore \sim 2\times 10^{11}\cm$ and then adiabatically cool as the gas expanded.  Instead, the photons must have been generated close to the radius where we observe them, $\rph \gtrsim {(\rm few)}\, \times 10^{13}\cm$.  We suspect that the expanding shell shocked on other gas, in order to make up for the adiabatic cooling as gas expanded from small radii, and to produce such a large amount of radiation relative to kinetic energy.
\end{enumerate}

\subsection{Implications \& questions for future work}
\label{sec:futurework}

\subsubsection{Energetics}
\citet{sq09} argued that many TDEs would lead to super-Eddington rates of BH feeding, and thus radiation-pressure-driven outflows would be likely.  They considered a case they called ``edge-dominated,'' in which the photosphere would be trapped in the sharp edge of the flow and would initially expand with time, rather than recede with time through a broadly-distributed wind (see Figure \ref{fig:rhotaushell} above).  The expanding shell we infer for PS1-10jh is similar to this picture, but with two significant differences: (1) \citet{sq09}  assumed that the outflow velocity would be $\sim 0.1c$ and so the region where photons would not be trapped (the diffusion time is less than the advection time) could be neglected; (2) \citet{sq09} did not expect the temperature to remain constant, although they did find only a weak time dependence for temperature $\teff \propto t^{-7/36}$, which could still be consistent with the data.

Our model of PS1-10jh as an expanding shell of gas leaves and raises several questions:
\begin{enumerate}
\item Why was the outflow so slow, having such tiny kinetic energy, only barely unbound from the BH (\S\ref{sec:energetics}, \S\ref{sec:interp_energy})?  Was the physics of the shock and accretion somehow regulated to distribute the radiation energy in such a way that the maximum amount of gas would be able to escape?  Could it be that the gas was given a distribution of velocities, but the fastest shells spread out so much that they quickly became optically thin (so invisible), and the photosphere always lay in the slowest (barely unbound) shell?
\item Why was the temperature close to constant, despite the fact that the photospheric radius ranged by an order of magnitude over the period of observation (Figure \ref{fig:roft})?  \citet{miller15} proposes that constant temperatures may be explained by disk winds: more super-Eddington feeding rates produce stronger disk winds, so that the accretion rate onto the BH is only weakly dependent on the feeding rate. Could this idea explain PS1-10jh's temperature in the context of the other physical properties of the event we have inferred?
\item Was the expansion of the shell non-adiabatic, as we argued in \S\ref{sec:adiabatic_cooling}?  In that section, we suggested that there may have been an additional shock, converting kinetic energy back to radiation.  What caused this shock? 
Could this process also explain the observation that the gas was barely unbound, and the roughly constant temperature?  
\end{enumerate}

We discuss these points further in the following subsection.

\subsubsection{Orbit circularization \& shocks at large radius}
\label{sec:extra_shock}
\citet{shiokawa15} perform general relativistic hydrodynamic simulations of the disruption of a white dwarf by a $500\msun$ BH, followed up by \citet{piran15}.  These authors find that the process of circularization of orbits of stellar gas particles following disruption is slower and less efficient than previously thought:  the shock at pericenter may be weak, and a second and third shock close to apocenter of the most tightly-bound gas may be partially responsible for circularization as well. They find that the time for (partial) circularization of orbits is $\sim 3 - 10$ times longer than previous expectations ($\sim (3-10)\tfall$: see eq. \ref{eq:tfall}), and the maximum feeding rate to the BH is thus $\sim 10$ times less.  Although these effects would stretch the black curve in our Figure \ref{fig:mdotomdotedd} for PS1-10jh out in time, they do not stretch it out enough to make it match the green curve:  the timescales for $\mdotfb$ and observed luminosity differ by a factor of $\sim 10^3$, not just a factor of $\sim 10$.

These authors indicate that the pericenter shock {\it is} strong if the disruption radius is close to the BH's event horizon (i.e., if $\rt \lesssim 5\rs$), since relativistic apsidal precession and potentially Lense-Thirring precession cause significant crossing of orbits \citep[e.g., recent work by][]{bonnerot15, hayasaki15, guillochon15}.  Provided $\mbh \gtrsim 10^5\msun$, the helium core disruption of PS1-10jh falls in the regime where the pericenter shock is strong enough to make circularization rapid.

\cite{piran15} suggest that the radiation seen in a tidal disruption event comes from the shocks near apocenter rather than the shock at pericenter (or from an accretion disk at yet smaller radii). However, we find that this does not work for PS1-10jh: the observed low mass ($\sim 0.01M_\odot$) combined with the low orbital velocity at $\rapo$ (see eq. \ref{eq:Rapo}) shows that the energy liberated by the apocenter shocks is
\begin{eqnarray}
E_{\rm apo} & \sim & \frac{2G \mbh M_{\rm shell}}{\rapo} \\
& \sim & 2 \times 10^{50}\left(\frac{\mbh}{2\times 10^5\msun}\right)\left(\frac{M_{\rm shell}}{0.01\msun}\right)\left(\frac{\rapo}{6\times 10^{12}\cm}\right)^{-1} \erg \, , \nonumber
\end{eqnarray}
an order of magnitude too small to explain the observed radiation energy.
Nevertheless, we found in \S\ref{sec:adiabatic_cooling} that the expanding shell likely experienced a shock at large radius that produced most of the observed radiation: we offer the speculation that perhaps gas that circularized through the apocenter shock was partially responsible (see \S\ref{sec:adiabatic_cooling}).
The process of circularizing the orbits of disrupted stellar material is clearly complicated and may be important for understanding PS1-10jh and other disruption candidates.  We defer further investigation of these ideas to future work.

\subsubsection{Photoionization calculations}

We reiterate that if ionized hydrogen had been present in significant quantities, it would have recombined in steady state to produce strong Balmer emission lines, which were not seen.  Recombination times were short (eq. \ref{eq:trec}); every hydrogen photoionization would have been balanced by a recombination cascade.  Arguments in the literature that ``over-ionization'' of hydrogen provides a means of suppressing Balmer emission do not make sense.  The only way to suppress Balmer emission is through absorption by excited {\it neutral} hydrogen, but velocity gradients in the outflowing gas suggest that H$\alpha$ would not have been optically thick, and so such absorption would not have been significant.  We conclude that PS1-10jh comprised gas that was helium-rich, not of solar composition. See Appendix \ref{sec:cloudy} for detailed explanation.

\subsubsection{Mechanism to produce helium core}

We argue in Appendix \ref{sec:cloudy} that the spectrum of PS1-10jh is inconsistent with the disruption of a main-sequence star, and argue instead for the disruption of the helium core of a stripped red giant star.  How could this come about?  Detailed investigation of this crucial question is beyond the scope of this work, but we briefly offer a few comments here.

First, could tidal stripping of a red giant star leave behind an object whose disruption could produce the requisite line ratio $\LHeHa$?  \citet{macleod12} study the tidal stripping by a BH of mass $\mbh =10^6\msun$ of a $1.4\msun$ star at various stages of post-main-sequence evolution.  Their RG I model has a core of mass $M_{\rm core} = 0.28\msun$ and envelope of mass $M_{\rm env} = 1.1\msun$.  The maximal stripping of the envelope (by the deepest disruption) removes $\sim 70\%$ of the envelope mass, leaving $\sim 0.3\msun$ still gravitationally bound to the core.  If we approximate that the envelope is purely hydrogen and the core is purely helium, and that the species fully mix following disruption, we find that this scenario leads to a hydrogen mass fraction $X = M_{\rm H}/(M_{\rm H}+M_{\rm He}) \sim 0.5$ (c.f. eq. \ref{eq:Xratio}), or a number density ratio
\begin{equation}\label{eq:stripped_ratio}
\frac{n_{\rm He}}{n_{\rm H}} \sim \frac{M_{\rm He}}{4M_{\rm H}} \sim 0.2 \, .
\end{equation}

Equation (\ref{eq:L4686Ha_simple}) shows that in the optically thin limit, $\LHeHa \propto n_{\rm He}/n_{\rm H}$.
Our Cloudy results in Figures \ref{fig:AGN_GMR_lineratio_linetau} and \ref{fig:BB3_5p5_lineratio_linetau} (top left parts of Panels (a) and (d)), which assume a density ratio $n_{\rm He}/n_{\rm H} = 0.1$, show that the line ratio under optically thin conditions\footnote{The higher values than in eq. (\ref{eq:L4686Ha_simple}) are due to the temperature dependence of the recombination coefficients.  We discuss the optically thin limit because velocity gradients in the outflowing gas likely reduce optical depths dramatically; see \S\ref{sec:cloudy_problems}.} is $\sim 0.4 - 1.8$.   This indicates that to reach the observed line ratio $\LHeHa > 5$ \citep{gezari12}, the number density ratio would need to be $\sim 3 - 10$ times as high as assumed by the Cloudy calculations, or $n_{\rm He}/n_{\rm H} \sim 0.3 - 1$. (If we make the same argument for \citealt{gaskell14}'s result of $\LHeHa > 3.7$, we find $n_{\rm He}/n_{\rm H} \sim 0.2 - 0.9$.)  On the low end of these results ($0.2 - 0.3$, which we derived for the incident SEDs closest to that of PS1-10jh), the number density ratios come close to those of \citet{macleod12}'s stripped red giants (eq. \ref{eq:stripped_ratio}).  This suggests that \citet{macleod12}'s stripping mechanism could be a viable one for producing a helium-rich core whose disruption spectrum would look like PS1-10jh's; complete stripping of the hydrogen envelope may not be required. Calculations by \citet{chengevans} also indicate that the outer layers of the (mostly) stripped core are tidally heated each time the core passes pericenter around the BH, and thus the remainder of the stellar envelope could be removed after several passes.

What happens after the envelope of the red giant is tidally stripped away?  \citet{bogdanovic_cheng} propose the following scenario.  Following the tidal stripping, half of the envelope is accreted and half escapes unbound, as would happen in full disruption of a solar-type star, though with an abrupt cut-off in accretion at late times because the core gravitationally retains the gas closest to it \citep{macleod12, guillochon13}.  (This accretion may have produced a bright flare, but evidently not during any period when this region of the sky was being observed.)  The stripped core then gradually becomes tidally heated by the BH's gravitational field, while its orbit slowly circularizes and decays by the emission of gravitational waves (as an EMRI: extreme mass-ratio inspiral).  For just the right orbital and stellar parameters, \citet{bogdanovic_cheng} find that the core will retain (rather than radiate away) its heat long enough to expand and ``lift its degeneracy'' enough to be tidally disrupted by the BH before gravitational waves carry it through the ISCO or event horizon.  Note that this scenario considers a BH of mass $\mbh \gtrsim 10^6\msun$, which is too massive to tidally disrupt a helium core that isn't tidally heated; a lower-mass BH ($\mbh \lesssim 2\times 10^5\msun$; eq. \ref{eq:rtcore}) can do the job without tidal heating.

In addition to hypothesizing scenarios to tidally disrupt helium-rich objects, it is also critical to consider how common such scenarios are likely to be.  \citet{macleod12} make careful estimates of disruption rates of stars at different stages of evolution (see their Figure 13 in particular).  Main-sequence stars are by far the most common disruptions, but red giant disruptions contribute $\sim 10\%$ to the total, especially for stars slightly more massive than the sun and above.  (This work did not consider BHs of low enough mass [$\mbh \lesssim 2\times 10^5\msun$] to disrupt red giant cores.)  Future research could combine and extend the work of \citet{macleod12} and \citet{bogdanovic_cheng} to determine theoretically how often the full scenario takes place: a red giant's core is tidally stripped of its hydrogen envelope, then it spirals in to the BH via gravitational wave emission, and then is ultimately disrupted.  The theoretical ratio of this rate to the rate of all disruptions would be illuminating for deciding how plausible this scenario is for explaining PS1-10jh.

\subsubsection{Comparison with other events}
As the number of tidal disruption candidates with detailed observations grows, preliminary comparisons of rates can become possible.  As mentioned above, PTF09ge \citep{kasliwal09, arcavi14} is a recently published candidate that also showed broad He II 4686 emission and no hydrogen.  Could this event also be the disruption of a stripped helium core?  Would the discovery of two such events be in line with theoretical rate expectations?  Other tidal disruption candidates with optical spectroscopy show both helium and hydrogen or emission, or only hydrogen emission \citep{arcavi14, holoien, vanvelzen11}, or no optical lines at all \citep{cenko11, vanvelzen11, chornock14}.

The type of analysis we have done for PS1-10jh required observations of the rise of the light curve, ultraviolet photometry, and spectroscopy, in order to derive the temperature and thus the photospheric radius, and to constrain the composition of the disrupted object.  We encourage transient surveys to seek all of these data for candidate disruptions:  it will be very useful to know if other events also show a clearly expanding then contracting photosphere, how fast it is moving, how massive is the gas, and the energetics of the event; and also to perform photoionization calculations to compare with spectra that can give clues to the composition of the disrupted object (and thus perhaps its density and constraints on the BH mass).

\subsubsection{Low-mass black hole}

We estimated that the mass of the BH was $\mbh \sim 2\times 10^5\msun$ or less, in order for the BH's tidal forces to be able to disrupt a dense core of a red giant outside the BH's ISCO.  (We independently estimated that $\mbh \sim 3 \times 10^5\msun$ by comparing the (half) linewidth of He II 4686 with the radius of the photosphere.)
Such modest BHs are expected to be very common in the universe, but are difficult to find via stellar dynamics or AGN emission, because the radius of influence and Eddington luminosities are so small.

PS1-10jh may offer an exciting first\footnote{Other events have claimed $\mbh$ measurements based on their light curves, but we believe that current understanding of the precise physical mechanisms producing the light curves is still too preliminary to make such claimed measurements.} significant BH mass constraint using a tidal disruption event.  The host galaxy of PS1-10jh has a total stellar mass around $3\times 10^9\msun$ but unknown morphology; SDSS contains an image of the galaxy, but with too poor resolution to decompose a bulge from a disk if present---so the bulge could be $\sim 10$ times less massive than the total stellar mass.  At these low masses, the $\mbh-M_{\rm bulge}$ relation increases in scatter and becomes poorly constrained \citep[e.g.,][]{greene08}. A BH of mass $2\times 10^5\msun$ in a host bulge of $3 \times (10^8 - 10^9)\msun$ appears roughly in line with previous $\mbh-M_{\rm bulge}$ results \citep[e.g.,][]{greene08, jiang11}. PS1-10jh could offer a new datapoint for the $\mbh-M_{\rm bulge}$ and $M-\sigma$ relations if additional observations (decomposing the bulge from the disk, and measuring the velocity dispersion) of the host are made.

\subsubsection{Super-Eddington accretion physics}

Our BH mass estimate $\mbh \sim 2\times 10^5\msun$ led us to conclude that the rate of feeding the BH was much higher than the Eddington rate, by a factor of $\sim 10^4$ close to the peak (eq. \ref{eq:mdotomdotedd}; Figure \ref{fig:mdotomdotedd}).

Such highly super-Eddington feeding generates a large amount of radiation and is expected to produce a complicated flow, with photon trapping, photon diffusion, outflows, accretion, and possibly other processes all playing roles \citep[e.g.,][]{ss, abramowicz, begelman79, Ohsuga_Mineshige07}.
Past studies of such flows have typically inferred low efficiencies for the conversion of mass feeding to escaping energy, because the gas becomes extremely optically thick and the diffusion of radiation out is ineffective; most of the radiation remains trapped and is soon swallowed by the BH along with the gas in which it is entrained.

 \citet{jiang14} recently performed a global 3-dimensional radiation magneto-hydrodynamic (MHD) simulation of super-Eddington accretion and a find significantly higher efficiency $\eta_{\rm esc} \equiv E_{\rm tot,out}/Mc^2 = 0.045$. Their simulation finds that the gas is turbulent, and cooling is dominated by magnetic buoyancy rather than photon diffusion; as a result, a significantly higher fraction of energy is able to escape.  They also find that the escaping energy is divided roughly $5:1$ between radiation and kinetic energy: in our terminology, $E_{\rm K} \sim 0.2 E_{\rm rad,out}$.  \citet{mckinney14} recently performed a 3D general-relativistic radiation MHD simulation of super-Eddington accretion onto a rapidly rotating BH.  They find that about 1\% of matter fed to the BH is converted to radiation ($\eta_{\rm rad} \sim 0.01$) and about 80\% of matter fed to the BH is actually accreted ($\eta_{\rm acc} \sim 1 - 0.8 = 0.2$ in their nomenclature).

It is interesting to compare these simulation results with accretion properties we infer for PS1-10jh.  The overall efficiency we find for PS1-10jh is $\eta_{\rm esc} \sim 0.02$  (\S\ref{sec:interp_energy}), similar to \citet{jiang14}'s result.
 Evidently, a remarkably high fraction of the mass of the disrupted star is converted to energy which manages to escape to great distances. However, we find that it does not escape immediately, but rather is trapped for $10^2 - 10^3$ times the inflow (fallback) time. \citet{jiang14} simulated a domain corresponding to size $r\ll\rapo$, so that their simulation could not account for any (potentially optically thick) post-apocenter shocked gas, as it is well outside the radius of their disk. We also infer a much lower kinetic energy ($E_{\rm K} \sim 5 \times 10^{-4} E_{\rm rad,out}$) than they do, but again, their simulation cannot address the effects of gas infalling from $\rapo$. They find that the mass-loss rate from the disk is $\sim30\%$ of the mass accretion rate onto the black hole, while we infer a ratio (on much larger scales) of ejected mass to accreted mass of $\sim 0.01\msun/0.15\msun \sim 7\%$.

Comparing with \citet{mckinney14}, we infer $\eta_{\rm rad} \sim 2E_{\rm rad,bol,out}/M_{\rm core}c^2 \sim 0.007$, close to their value of 0.01 (though ours is integrated over time, rather than comparing at a specific moment). They find that their wind carries away some 20\% of the mass, midway between the result of \citet{jiang14} and our inferred ratio for PS1-10jh.

Future super-Eddington accretion simulations may focus more on tidal disruptions by starting from initial conditions given by studies of orbital circularization of falling-back stellar debris (see references in \S\ref{sec:extra_shock}).
Continued comparisons between observed properties of tidal disruption candidates and accretion simulations may give further insights into accretion physics relevant in other contexts as well, such as ultra-luminous X-ray sources and the growth of massive BHs at high redshift.

\section*{Acknowledgments}
We would like to thank Eliot Quataert, Brad Cenko, David Arnett, Shane Davis, Suvi Gezari, Tamara Bogdanovi{\'c}, James Owen, Nick Stone, Cole Miller, Natalie Price-Jones, Shelley Wright, Anil Seth, Richard Shaw, David Strubbe, and Hugo Strubbe for helpful discussions.  We thank Avishay Gal-Yam for permission to adapt his Figure 1 from \citet{gal-yam12} for our Figure \ref{fig:GalYamcomp}. LES acknowledges a CITA Postdoctoral Fellowship (2012-2015) and a Science Teaching \& Learning Fellowship (2015-) from the UBC Department of Physics \& Astronomy / Carl Wieman Science Education Initiative.

\bibliography{Strubbe_Murray}

\begin{thebibliography}{73}
\expandafter\ifx\csname natexlab\endcsname\relax\def\natexlab#1{#1}\fi

\bibitem[{{Abbott}(1982)}]{abbott82}
{Abbott} D.~C., 1982, \apj, 259, 282

\bibitem[{{Abramowicz} {et~al}\mbox{.}(1988){Abramowicz}, {Czerny}, {Lasota},
  \& {Szuszkiewicz}}]{abramowicz}
{Abramowicz} M.~A., {Czerny} B., {Lasota} J.~P., {Szuszkiewicz} E., 1988, \apj,
  332, 646

\bibitem[{{Aihara} {et~al}\mbox{.}(2011){Aihara}, {Allende Prieto}, {An},
  {Anderson}, {Aubourg}, {Balbinot}, {Beers}, {Berlind}, {Bickerton},
  {Bizyaev}, {Blanton}, {Bochanski}, {Bolton}, {Bovy}, {Brandt}, {Brinkmann},
  {Brown}, {Brownstein}, {Busca}, {Campbell}, {Carr}, {Chen}, {Chiappini},
  {Comparat}, {Connolly}, {Cortes}, {Croft}, {Cuesta}, {da Costa}, {Davenport},
  {Dawson}, {Dhital}, {Ealet}, {Ebelke}, {Edmondson}, {Eisenstein},
  {Escoffier}, {Esposito}, {Evans}, {Fan}, {Femen{\'{\i}}a Castell{\'a}},
  {Font-Ribera}, {Frinchaboy}, {Ge}, {Gillespie}, {Gilmore}, {Gonz{\'a}lez
  Hern{\'a}ndez}, {Gott}, {Gould}, {Grebel}, {Gunn}, {Hamilton}, {Harding},
  {Harris}, {Hawley}, {Hearty}, {Ho}, {Hogg}, {Holtzman}, {Honscheid}, {Inada},
  {Ivans}, {Jiang}, {Johnson}, {Jordan}, {Jordan}, {Kazin}, {Kirkby}, {Klaene},
  {Knapp}, {Kneib}, {Kochanek}, {Koesterke}, {Kollmeier}, {Kron}, {Lampeitl},
  {Lang}, {Le Goff}, {Lee}, {Lin}, {Long}, {Loomis}, {Lucatello}, {Lundgren},
  {Lupton}, {Ma}, {MacDonald}, {Mahadevan}, {Maia}, {Makler}, {Malanushenko},
  {Malanushenko}, {Mandelbaum}, {Maraston}, {Margala}, {Masters}, {McBride},
  {McGehee}, {McGreer}, {M{\'e}nard}, {Miralda-Escud{\'e}}, {Morrison},
  {Mullally}, {Muna}, {Munn}, {Murayama}, {Myers}, {Naugle}, {Neto}, {Nguyen},
  {Nichol}, {O'Connell}, {Ogando}, {Olmstead}, {Oravetz}, {Padmanabhan},
  {Palanque-Delabrouille}, {Pan}, {Pandey}, {P{\^a}ris}, {Percival},
  {Petitjean}, {Pfaffenberger}, {Pforr}, {Phleps}, {Pichon}, {Pieri}, {Prada},
  {Price-Whelan}, {Raddick}, {Ramos}, {Reyl{\'e}}, {Rich}, {Richards}, {Rix},
  {Robin}, {Rocha-Pinto}, {Rockosi}, {Roe}, {Rollinde}, {Ross}, {Ross},
  {Rossetto}, {S{\'a}nchez}, {Sayres}, {Schlegel}, {Schlesinger}, {Schmidt},
  {Schneider}, {Sheldon}, {Shu}, {Simmerer}, {Simmons}, {Sivarani}, {Snedden},
  {Sobeck}, {Steinmetz}, {Strauss}, {Szalay}, {Tanaka}, {Thakar}, {Thomas},
  {Tinker}, {Tofflemire}, {Tojeiro}, {Tremonti}, {Vandenberg}, {Vargas
  Maga{\~n}a}, {Verde}, {Vogt}, {Wake}, {Wang}, {Weaver}, {Weinberg}, {White},
  {White}, {Yanny}, {Yasuda}, {Yeche}, \& {Zehavi}}]{SDSS}
{Aihara} H. {et~al.}, 2011, \apjs, 193, 29

\bibitem[{{Arcavi} {et~al}\mbox{.}(2014){Arcavi}, {Gal-Yam}, {Sullivan}, {Pan},
  {Cenko}, {Horesh}, {Ofek}, {De Cia}, {Yan}, {Yang}, {Howell}, {Tal},
  {Kulkarni}, {Tendulkar}, {Tang}, {Xu}, {Sternberg}, {Cohen}, {Bloom},
  {Nugent}, {Kasliwal}, {Perley}, {Quimby}, {Miller}, {Theissen}, \&
  {Laher}}]{arcavi14}
{Arcavi} I. {et~al.}, 2014, \apj, 793, 38

\bibitem[{{Armijo} \& {de Freitas Pacheco}(2013)}]{armijo13}
{Armijo} M.~M., {de Freitas Pacheco} J.~A., 2013, \mnras, 430, L45

\bibitem[{{Arnett}(1982)}]{arnett82}
{Arnett} W.~D., 1982, \apj, 253, 785

\bibitem[{{Begelman}(1979)}]{begelman79}
{Begelman} M.~C., 1979, \mnras, 187, 237

\bibitem[{Bloom {et~al}\mbox{.}(2011)Bloom, Giannios, Metzger, Cenko, Perley,
  Butler, Tanvir, Levan, O'~Brien, Strubbe, De~Colle, Ramirez-Ruiz, Lee,
  Nayakshin, Quataert, King, Cucchiara, Guillochon, Bower, Fruchter, Morgan, \&
  van~der Horst}]{bloom11}
Bloom J.~S. {et~al.}, 2011, Science, 333, 203

\bibitem[{{Bogdanovi{\'c}}, {Cheng} \& {Amaro-Seoane}(2014){Bogdanovi{\'c}},
  {Cheng}, \& {Amaro-Seoane}}]{bogdanovic_cheng}
{Bogdanovi{\'c}} T., {Cheng} R.~M., {Amaro-Seoane} P., 2014, \apj, 788, 99

\bibitem[{{Bonnerot} {et~al}\mbox{.}(2015){Bonnerot}, {Rossi}, {Lodato}, \&
  {Price}}]{bonnerot15}
{Bonnerot} C., {Rossi} E.~M., {Lodato} G., {Price} D.~J., 2015, ArXiv e-prints

\bibitem[{{Castor}, {Abbott} \& {Klein}(1975){Castor}, {Abbott}, \&
  {Klein}}]{castor75}
{Castor} J.~I., {Abbott} D.~C., {Klein} R.~I., 1975, \apj, 195, 157

\bibitem[{{Cenko} {et~al}\mbox{.}(2012){Cenko}, {Bloom}, {Kulkarni}, {Strubbe},
  {Miller}, {Butler}, {Quimby}, {Gal-Yam}, {Ofek}, {Quataert}, {Bildsten},
  {Poznanski}, {Perley}, {Morgan}, {Filippenko}, {Frail}, {Arcavi}, {Ben-Ami},
  {Cucchiara}, {Fassnacht}, {Green}, {Hook}, {Howell}, {Lagattuta}, {Law},
  {Kasliwal}, {Nugent}, {Silverman}, {Sullivan}, {Tendulkar}, \&
  {Yaron}}]{cenko11}
{Cenko} S.~B. {et~al.}, 2012, \mnras, 420, 2684

\bibitem[{{Cheng} \& {Evans}(2013)}]{chengevans}
{Cheng} R.~M., {Evans} C.~R., 2013, \prd, 87, 104010

\bibitem[{{Chornock} {et~al}\mbox{.}(2014){Chornock}, {Berger}, {Gezari},
  {Zauderer}, {Rest}, {Chomiuk}, {Kamble}, {Soderberg}, {Czekala}, {Dittmann},
  {Drout}, {Foley}, {Fong}, {Huber}, {Kirshner}, {Lawrence}, {Lunnan},
  {Marion}, {Narayan}, {Riess}, {Roth}, {Sanders}, {Scolnic}, {Smartt},
  {Smith}, {Stubbs}, {Tonry}, {Burgett}, {Chambers}, {Flewelling}, {Hodapp},
  {Kaiser}, {Magnier}, {Martin}, {Neill}, {Price}, \& {Wainscoat}}]{chornock14}
{Chornock} R. {et~al.}, 2014, \apj, 780, 44

\bibitem[{{Davies} \& {King}(2005)}]{davies05}
{Davies} M.~B., {King} A., 2005, \apjl, 624, L25

\bibitem[{{Evans} \& {Kochanek}(1989)}]{evkoch}
{Evans} C.~R., {Kochanek} C.~S., 1989, \apjl, 346, L13

\bibitem[{{Ferland} {et~al}\mbox{.}(1992){Ferland}, {Peterson}, {Horne},
  {Welsh}, \& {Nahar}}]{ferland92}
{Ferland} G.~J., {Peterson} B.~M., {Horne} K., {Welsh} W.~F., {Nahar} S.~N.,
  1992, \apj, 387, 95

\bibitem[{{Ferland} {et~al}\mbox{.}(2013){Ferland}, {Porter}, {van Hoof},
  {Williams}, {Abel}, {Lykins}, {Shaw}, {Henney}, \& {Stancil}}]{cloudy13}
{Ferland} G.~J. {et~al.}, 2013, \rmxaa, 49, 137

\bibitem[{{Gal-Yam}(2012)}]{gal-yam12}
{Gal-Yam} A., 2012, Science, 337, 927

\bibitem[{{Gaskell} \& {Rojas Lobos}(2014)}]{gaskell14}
{Gaskell} C.~M., {Rojas Lobos} P.~A., 2014, \mnras, 438, L36

\bibitem[{{Gezari} {et~al}\mbox{.}(2012){Gezari}, {Chornock}, {Rest},
  {et~al.}}]{gezari12}
{Gezari} S., {Chornock} R., {Rest} A., {et~al.}, 2012, \nat, 485, 217

\bibitem[{{Gezari} {et~al}\mbox{.}(2009){Gezari}, {Heckman}, {Cenko},
  {et~al.}}]{gezari09}
{Gezari} S., {Heckman} T., {Cenko} S.~B., {et~al.}, 2009, \apj, 698, 1367

\bibitem[{{Greene}, {Ho} \& {Barth}(2008){Greene}, {Ho}, \& {Barth}}]{greene08}
{Greene} J.~E., {Ho} L.~C., {Barth} A.~J., 2008, \apj, 688, 159

\bibitem[{{Guillochon}, {Manukian} \& {Ramirez-Ruiz}(2014){Guillochon},
  {Manukian}, \& {Ramirez-Ruiz}}]{guillochon14}
{Guillochon} J., {Manukian} H., {Ramirez-Ruiz} E., 2014, \apj, 783, 23

\bibitem[{{Guillochon} \& {Ramirez-Ruiz}(2013)}]{guillochon13}
{Guillochon} J., {Ramirez-Ruiz} E., 2013, \apj, 767, 25

\bibitem[{{Guillochon} \& {Ramirez-Ruiz}(2015)}]{guillochon15}
{Guillochon} J., {Ramirez-Ruiz} E., 2015, ArXiv e-prints

\bibitem[{{Guillochon} {et~al}\mbox{.}(2009){Guillochon}, {Ramirez-Ruiz},
  {Rosswog}, \& {Kasen}}]{guillochon09}
{Guillochon} J., {Ramirez-Ruiz} E., {Rosswog} S., {Kasen} D., 2009, \apj, 705,
  844

\bibitem[{{Haas} {et~al}\mbox{.}(2012){Haas}, {Shcherbakov}, {Bode}, \&
  {Laguna}}]{haas}
{Haas} R., {Shcherbakov} R.~V., {Bode} T., {Laguna} P., 2012, \apj, 749, 117

\bibitem[{{H{\"a}ring} \& {Rix}(2004)}]{haring04}
{H{\"a}ring} N., {Rix} H., 2004, \apjl, 604, L89

\bibitem[{{Hayasaki}, {Stone} \& {Loeb}(2015){Hayasaki}, {Stone}, \&
  {Loeb}}]{hayasaki15}
{Hayasaki} K., {Stone} N.~C., {Loeb} A., 2015, ArXiv e-prints

\bibitem[{{Holoien} {et~al}\mbox{.}(2014){Holoien}, {Prieto}, {Bersier},
  {Kochanek}, {Stanek}, {Shappee}, {Grupe}, {Basu}, {Beacom}, {Brimacombe},
  {Brown}, {Davis}, {Jencson}, {Pojmanski}, \& {Szczygie{\l}}}]{holoien}
{Holoien} T.~W.-S. {et~al.}, 2014, \mnras, 445, 3263

\bibitem[{{Howarth} \& {Prinja}(1989)}]{howarth89}
{Howarth} I.~D., {Prinja} R.~K., 1989, \apjs, 69, 527

\bibitem[{{Iglesias} \& {Rogers}(1996)}]{opal}
{Iglesias} C.~A., {Rogers} F.~J., 1996, \apj, 464, 943

\bibitem[{{Jiang}, {Greene} \& {Ho}(2011){Jiang}, {Greene}, \& {Ho}}]{jiang11}
{Jiang} Y.-F., {Greene} J.~E., {Ho} L.~C., 2011, \apjl, 737, L45

\bibitem[{{Jiang}, {Stone} \& {Davis}(2014){Jiang}, {Stone}, \&
  {Davis}}]{jiang14}
{Jiang} Y.-F., {Stone} J.~M., {Davis} S.~W., 2014, \apj, 796, 106

\bibitem[{{Kaiser} {et~al}\mbox{.}(2010){Kaiser}, {Burgett}, {Chambers},
  {Denneau}, {Heasley}, {Jedicke}, {Magnier}, {Morgan}, {Onaka}, \&
  {Tonry}}]{kaiser10}
{Kaiser} N. {et~al.}, 2010, in Society of Photo-Optical Instrumentation
  Engineers (SPIE) Conference Series, Vol. 7733, Society of Photo-Optical
  Instrumentation Engineers (SPIE) Conference Series, p.~0

\bibitem[{{Kasliwal} {et~al}\mbox{.}(2009){Kasliwal}, {Kulkarni}, {Quimby},
  {Nugent}, {Howell}, {Cooke}, {Cenko}, {Gal-Yam}, {Law}, {Levitan}, {Ofek}, \&
  {Poznanski}}]{kasliwal09}
{Kasliwal} M.~M. {et~al.}, 2009, The Astronomer's Telegram, 2055, 1

\bibitem[{{Kawabata} {et~al}\mbox{.}(2010){Kawabata}, {Maeda}, {Nomoto},
  {Taubenberger}, {Tanaka}, {Deng}, {Pian}, {Hattori}, \& {Itagaki}}]{kawabata}
{Kawabata} K.~S. {et~al.}, 2010, \nat, 465, 326

\bibitem[{{Kesden}(2012)}]{kesden}
{Kesden} M., 2012, \prd, 85, 024037

\bibitem[{{Kochanek}(1994)}]{kochanek94}
{Kochanek} C.~S., 1994, \apj, 422, 508

\bibitem[{{Komossa}(2002)}]{komossa02}
{Komossa} S., 2002, in Rev. Mod. Ast., Vol.~15, JENAM 2001: Astronomy with
  Large Telescopes from Ground and Space, {R.~E.~Schielicke}, ed., p.~27

\bibitem[{{Korista} \& {Goad}(2000)}]{koristagoad2000}
{Korista} K.~T., {Goad} M.~R., 2000, \apj, 536, 284

\bibitem[{{Korista} \& {Goad}(2004)}]{koristagoad}
{Korista} K.~T., {Goad} M.~R., 2004, \apj, 606, 749

\bibitem[{{Lacy}, {Townes} \& {Hollenbach}(1982){Lacy}, {Townes}, \&
  {Hollenbach}}]{lacy}
{Lacy} J.~H., {Townes} C.~H., {Hollenbach} D.~J., 1982, \apj, 262, 120

\bibitem[{{Lamers} \& {Cassinelli}(1999)}]{lamers99}
{Lamers} H.~J.~G.~L.~M., {Cassinelli} J.~P., 1999, {Introduction to Stellar
  Winds}

\bibitem[{{Lawrence} {et~al}\mbox{.}(2007){Lawrence}, {Warren}, {Almaini},
  {Edge}, {Hambly}, {Jameson}, {Lucas}, {Casali}, {Adamson}, {Dye}, {Emerson},
  {Foucaud}, {Hewett}, {Hirst}, {Hodgkin}, {Irwin}, {Lodieu}, {McMahon},
  {Simpson}, {Smail}, {Mortlock}, \& {Folger}}]{ukidss}
{Lawrence} A. {et~al.}, 2007, \mnras, 379, 1599

\bibitem[{{Li}, {Narayan} \& {Menou}(2002){Li}, {Narayan}, \& {Menou}}]{li02}
{Li} L., {Narayan} R., {Menou} K., 2002, \apj, 576, 753

\bibitem[{{Lodato}, {King} \& {Pringle}(2009){Lodato}, {King}, \&
  {Pringle}}]{lodato09}
{Lodato} G., {King} A.~R., {Pringle} J.~E., 2009, \mnras, 392, 332

\bibitem[{{MacLeod}, {Guillochon} \& {Ramirez-Ruiz}(2012){MacLeod},
  {Guillochon}, \& {Ramirez-Ruiz}}]{macleod12}
{MacLeod} M., {Guillochon} J., {Ramirez-Ruiz} E., 2012, \apj, 757, 134

\bibitem[{{Martin} {et~al}\mbox{.}(2005){Martin}, {Fanson}, {Schiminovich},
  {Morrissey}, {Friedman}, {Barlow}, {Conrow}, {Grange}, {Jelinsky},
  {Milliard}, {Siegmund}, {Bianchi}, {Byun}, {Donas}, {Forster}, {Heckman},
  {Lee}, {Madore}, {Malina}, {Neff}, {Rich}, {Small}, {Surber}, {Szalay},
  {Welsh}, \& {Wyder}}]{martin05}
{Martin} D.~C. {et~al.}, 2005, \apjl, 619, L1

\bibitem[{{Maxted} {et~al}\mbox{.}(2011){Maxted}, {Anderson}, {Burleigh},
  {Collier Cameron}, {Heber}, {G{\"a}nsicke}, {Geier}, {Kupfer}, {Marsh},
  {Nelemans}, {O'Toole}, {{\O}stensen}, {Smalley}, \& {West}}]{maxted11}
{Maxted} P.~F.~L. {et~al.}, 2011, \mnras, 418, 1156

\bibitem[{{McKinney} {et~al}\mbox{.}(2014){McKinney}, {Tchekhovskoy},
  {Sadowski}, \& {Narayan}}]{mckinney14}
{McKinney} J.~C., {Tchekhovskoy} A., {Sadowski} A., {Narayan} R., 2014, \mnras,
  441, 3177

\bibitem[{{Miller}(2015)}]{miller15}
{Miller} M.~C., 2015, ArXiv e-prints

\bibitem[{{Murray} {et~al}\mbox{.}(1995){Murray}, {Chiang}, {Grossman}, \&
  {Voit}}]{murray95}
{Murray} N., {Chiang} J., {Grossman} S.~A., {Voit} G.~M., 1995, \apj, 451, 498

\bibitem[{{Ohsuga} \& {Mineshige}(2007)}]{Ohsuga_Mineshige07}
{Ohsuga} K., {Mineshige} S., 2007, \apj, 670, 1283

\bibitem[{{Osterbrock}(1989)}]{osterbrock}
{Osterbrock} D.~E., 1989, {Astrophysics of gaseous nebulae and active galactic
  nuclei}

\bibitem[{{Paxton} {et~al}\mbox{.}(2011){Paxton}, {Bildsten}, {Dotter},
  {Herwig}, {Lesaffre}, \& {Timmes}}]{paxtonmesa}
{Paxton} B., {Bildsten} L., {Dotter} A., {Herwig} F., {Lesaffre} P., {Timmes}
  F., 2011, \apjs, 192, 3

\bibitem[{{Phinney}(1989)}]{phinney89}
{Phinney} E.~S., 1989, in IAU Symposium, Vol. 136, The Center of the Galaxy,
  {M.~Morris}, ed., p. 543

\bibitem[{{Piran} {et~al}\mbox{.}(2015){Piran}, {Svirski}, {Krolik}, {Cheng},
  \& {Shiokawa}}]{piran15}
{Piran} T., {Svirski} G., {Krolik} J., {Cheng} R.~M., {Shiokawa} H., 2015,
  ArXiv e-prints

\bibitem[{{Proga}, {Stone} \& {Kallman}(2000){Proga}, {Stone}, \&
  {Kallman}}]{proga00}
{Proga} D., {Stone} J.~M., {Kallman} T.~R., 2000, \apj, 543, 686

\bibitem[{{Quimby} {et~al}\mbox{.}(2011){Quimby}, {Kulkarni}, {Kasliwal},
  {Gal-Yam}, {Arcavi}, {Sullivan}, {Nugent}, {Thomas}, {Howell}, {Nakar},
  {Bildsten}, {Theissen}, {Law}, {Dekany}, {Rahmer}, {Hale}, {Smith}, {Ofek},
  {Zolkower}, {Velur}, {Walters}, {Henning}, {Bui}, {McKenna}, {Poznanski},
  {Cenko}, \& {Levitan}}]{quimby}
{Quimby} R.~M. {et~al.}, 2011, \nat, 474, 487

\bibitem[{{Rees}(1988)}]{rees}
{Rees} M.~J., 1988, \nat, 333, 523

\bibitem[{{Rees} \& {Meszaros}(1994)}]{grb_internal}
{Rees} M.~J., {Meszaros} P., 1994, \apjl, 430, L93

\bibitem[{{Rossi} \& {Begelman}(2009)}]{rossi09}
{Rossi} E.~M., {Begelman} M.~C., 2009, \mnras, 392, 1451

\bibitem[{{Shakura} \& {Sunyaev}(1973)}]{ss}
{Shakura} N.~I., {Sunyaev} R.~A., 1973, \aap, 24, 337

\bibitem[{{Shiokawa} {et~al}\mbox{.}(2015){Shiokawa}, {Krolik}, {Cheng},
  {Piran}, \& {Noble}}]{shiokawa15}
{Shiokawa} H., {Krolik} J.~H., {Cheng} R.~M., {Piran} T., {Noble} S.~C., 2015,
  ArXiv e-prints

\bibitem[{{Smartt}(2009)}]{2009ARA&A..47...63S}
{Smartt} S.~J., 2009, \araa, 47, 63

\bibitem[{{Smith} {et~al}\mbox{.}(2008){Smith}, {Chornock}, {Li},
  {Ganeshalingam}, {Silverman}, {Foley}, {Filippenko}, \& {Barth}}]{smith08}
{Smith} N., {Chornock} R., {Li} W., {Ganeshalingam} M., {Silverman} J.~M.,
  {Foley} R.~J., {Filippenko} A.~V., {Barth} A.~J., 2008, \apj, 686, 467

\bibitem[{{Strubbe} \& {Quataert}(2009)}]{sq09}
{Strubbe} L.~E., {Quataert} E., 2009, \mnras, 400, 2070

\bibitem[{{Strubbe} \& {Quataert}(2011)}]{sq11}
{Strubbe} L.~E., {Quataert} E., 2011, \mnras, 415, 168

\bibitem[{{van Velzen} {et~al}\mbox{.}(2011){van Velzen}, {Farrar}, {Gezari},
  {et~al.}}]{vanvelzen11}
{van Velzen} S., {Farrar} G.~R., {Gezari} S., {et~al.}, 2011, \apj, 741, 73

\bibitem[{{van Velzen} {et~al}\mbox{.}(2013){van Velzen}, {Frail},
  {K{\"o}rding}, \& {Falcke}}]{vanvelzen13}
{van Velzen} S., {Frail} D.~A., {K{\"o}rding} E., {Falcke} H., 2013, \aap, 552,
  A5

\bibitem[{{Verner} \& {Ferland}(1996)}]{vernerferland}
{Verner} D.~A., {Ferland} G.~J., 1996, \apjs, 103, 467

\end{thebibliography}

\appendix
\section{Photoionization calculations:  Not the disruption of a main-sequence star}
\label{sec:cloudy}
In this section we show that the observed spectrum of PS1-10jh is unlikely to be produced by the disruption of a main-sequence star.  We perform calculations using the publicly available photoionization code Cloudy \citep[last described by][]{cloudy13}, and compare calculated line ratios of He II 4686 to H$\alpha$ with the lower limit observed for PS1-10jh. \citet{gezari12} find that any H$\alpha$ emission is at least 5 times fainter than the He II $4686\ang$ emission\footnote{\citet{gezari12} published no error bars for this number, because of the difficulty in subtracting off the unknown brightness of the host galaxy.  Now that the event has faded below the brightness of the host, it is possible to determine more precise constraints on $\LHeHa$, as Gezari et al. are currently working on (personal communication).} ($\LHeHa > 5$); \citet{gaskell14} separately analyze \citet{gezari12}'s spectrum and suggest that $\LHeHa > 3.7 \pm 25\%$.

\subsection{Review of photoionization physics}
\label{sec:photoion_review}
We begin by briefly reviewing the physics of photoionization and how abundance limits may be inferred from observed line ratios.  Photons incident from a central source (perhaps an accretion disk close to the BH's event horizon) strike a cloud of gas some distance away from the source; if the photons are energetic enough, they photoionize atoms and ions in the cloud.  After a recombination time $t_{\rm rec}$ (eq. \ref{eq:trec}), electrons and ions begin colliding and recombining, generally to an excited state of the atom or ion; the electron then spontaneously falls to lower and lower energy levels, emitting line photons with each transition.  In an optically thin region (from which all line photons can escape), the rate of emission of a given line transition is
\begin{equation}
L_{\rm line} \sim n_{\rm i}n_{\rm e}\alpha^{\rm line}_{\rm rec}(n, T)V (h\nu_{\rm line}) \, ,
\end{equation}
where $n_{\rm i}$ and $n_{\rm e}$ are respectively the ion and electron number densities, $\alpha^{\rm line}_{\rm rec}$ is the effective recombination coefficient of the transition, $V$ is the emitting volume, and $h\nu_{\rm line}$ is the energy of the transition. He II $4686\ang$ and H$\alpha$ are lines of singly ionized helium ($4\rightarrow 3)$ and neutral hydrogen ($3\rightarrow 2)$, with well-known recombination coefficients. For example, at $T = 2 \times 10^4\K$,
$\alpha_{4686} = 1.77 \times 10^{-13}\cm^3\s^{-1}$ is the effective Case B recombination coefficient for He II 4686, and $\alpha_{\rm H\alpha} = 6.04 \times 10^{-14}\cm^3\s^{-1}$ is the effective Case B recombination coefficient\footnote{\citet{osterbrock} gives $\alpha_{\rm H\beta} = 1.62 \times 10^{-14}\cm^3\s^{-1}$ and $j_{\rm H\alpha}/j_{\rm H\beta} = 2.76$, both at $T = 2\times 10^4\K$, where $j_{\rm Hx} \propto \alpha_{\rm Hx}\nu_{\rm Hx}$.} for H$\alpha$ \citep[e.g.,][]{osterbrock}.  (Note that the ratio of these coefficients is 2.9; more discussion of the temperature dependence of this ratio is below in \S\ref{sec:cloudy_interp}.)
The line flux ratio of He II 4686 to H$\alpha$ in an optically thin region is therefore approximately
\begin{eqnarray} \label{eq:L4686Ha_simple}
\frac{L_{4686}}{L_{\rm H\alpha}} & \approx & \frac{n_{\rm e} \nhepp\, \alpha_{4686}\, (h\nu_{4686})}{n_{\rm e}n_{\rm H^+}\, \alpha_{\rm H\alpha}\, (h\nu_{\rm H\alpha})} \nonumber \\
& \approx & 0.4 \left(\frac{\nhepp/n_{\rm H^+}}{0.1}\right) \left(\frac{\alpha_{4686} / \alpha_{\rm H\alpha}}{2.9}\right) \, .
\end{eqnarray}

If the following conditions are met---(1) hydrogen and helium are both highly ionized (so that $n_{\rm H} \approx n_{\rm H^{+}}$ and $n_{\rm He} \approx n_{\rm He^{++}}$), (2) the region is optically thin to H$\alpha$ and $4686\ang$ photons, and (3) the temperature of the gas is relatively low ($T\ll 10^6\K$ so that $\alpha_{4686}/\alpha_{\rm H\alpha} \approx 2.9$)---then the line ratio offers a direct estimate of the abundance ratio of helium to hydrogen.  The line ratio lower limit $\LHeHa > 5$ implies $n_{\rm He}/n_{\rm H} > 1.2$, which corresponds to a mass fraction
\begin{equation}\label{eq:Xratio}
X = \frac{n_{\rm H}}{n_{\rm H}+4n_{\rm He}} < 0.2 \, ,
\end{equation}
far below the abundance ratio $X \approx 0.7$ found in solar-type stars.
\citet{gezari12} make essentially this argument\footnote{In \citet{gezari12} Supplementary Information \S7; we have fixed the small mistake where He$^+$ and H$^0$ were written instead of He$^{++}$ and H$^+$.} and propose on this basis that PS1-10jh is the disruption of a helium-rich object, such as the helium core of a stripped red giant star.

These simplifying conditions are generally found for relatively high fluxes (to highly ionize the gas), low densities (to keep lines optically thin) and soft SEDs (to keep the gas relatively cool).  When these simplifying conditions are not met, other additional effects become important, e.g., collisional ionization, 3-body recombination, and line absorption.  Numerical calculations, such as those using Cloudy, become crucial for capturing these complicated radiative effects.

\subsection{Proposal that PS1-10jh is disruption of a main-sequence star}

\citet{guillochon14} propose that PS1-10jh is the disruption of a main sequence star: they propose that hydrogen is indeed present in cosmic abundances in the outflowing material, but radiative transfer effects suppress emission from hydrogen relative to helium below detectable limits.  They appeal to Figure 1 in \citet{koristagoad}, who use Cloudy to calculate theoretical equivalent widths\footnote{These equivalent widths are relative to the continuum flux at $1215\ang$.} (EWs) of various emission lines in the spectra of active galactic nuclei (AGN).  The Cloudy calculations assume cosmic abundances of hydrogen and helium ($n_{\rm He} = 0.1n_{\rm H}$), and are performed for different combinations of hydrogen gas density $n_{\rm H}$ and hydrogen-ionizing number flux
\begin{equation}
\Phi\equiv \frac{1}{4\pi r^2} \int_{\chi_{\rm H}}^{\infty}\frac{L_\nu}{h\nu}\, d\nu \, ,
\end{equation}
with $\chi_{\rm H} = 13.6\eV$ the ionization potential of hydrogen.
In the region of this figure where $n_{\rm H}\sim 10^{12}\cm^{-3}$ and $\Phi\sim 10^{22} \s^{-1}\cm^{-2}$, it is seen that the He II 4686 EW is a few $\rm{\AA}$, while the H$\alpha$ EW is $< 1\rm{\AA}$ (although it is not clear from the figure that He II 4686 is fully 5 times brighter than H$\alpha$ even in this region).  Guillochon et al.'s idea is that PS1-10jh was in this region of parameter space during the two epochs when spectra were observed: they assert that the lack of observed H$\alpha$ is consistent with \citet{koristagoad}'s calculated line ratios.  Since \citet{koristagoad}'s calculations assume cosmic abundances of hydrogen and helium, as would be present in a main sequence star, Guillochon et al. assert that PS1-10jh is the tidal disruption of a main sequence star.

However, \citet{guillochon14} draw these conclusions without taking into account the magnitude of the observed lower limit of He II 4686 to H$\alpha$ emission ($\LHeHa > 5$), and without performing their own Cloudy photoionization calculations for the relevant input parameters.
\citet{gaskell14} do perform Cloudy calculations to investigate the line ratio for PS1-10jh, and find that the ratio $\LHeHa$ can be $\sim 4$ for a narrow range in density around $\sim 10^{11}\cm^{-3}$. They explain that this occurs when H$\alpha$ becomes optically thick and the line ratio approaches its blackbody limit (see below our eq. \ref{eq:BB_line_ratio} and surrounding discussion).  However, as explained further below, (1) these Cloudy calculations assume a hard AGN spectral energy distribution when a thermal SED is more appropriate, and (2) it is likely that these calculations are not converged.  Previous work has also not considered how velocity gradients in the gas dramatically reduce line optical depths and likely reduce maximal line ratios, which is likely a serious problem for all of these results so far (see \S\ref{sec:cloudy_problems}).

We study these issues in detail, and find that there are significant problems with the interpretation that PS1-10jh is the disruption of a main sequence star.

\subsection{Cloudy calculations}
\label{sec:cloudy_calcs}

We have performed suites of photoionization calculations using the latest release of Cloudy C13 (version C13.01\footnote{Most recent Cloudy version C13.03 has no revisions strongly relevant to our calculations.}) \citep{cloudy13}, looking for regions of parameter space where the He II 4686 line luminosity is more than 5 times the H$\alpha$ line luminosity.
Note that Figure 1 in \citet{koristagoad} was created from calculations using significantly older Cloudy version v90.04.  The updated version of Cloudy includes a much more sophisticated treatment of the hydrogenic and helium-like atomic energy levels, giving more accurate line luminosities for transitions between these energy levels (K. Korista, personal communication), crucial for investigating $\LHeHa$ in PS1-10jh.  Unfortunately, however, in versions to date, Cloudy has difficulty converging as the number of ``resolved'' atomic energy levels increases, in the region of parameter space where both the density and incident flux are high (R. Porter, personal communication).  Although not discussed in \citet{guillochon14} and \citet{gaskell14}, these issues should present problems for conclusions in these papers.  We present our Cloudy results with caveats about convergence where relevant.

An important consideration in Cloudy calculations is the spectral energy distribution (SED) of the source whose light is incident on the theoretical cloud of gas whose emission lines we wish to model.  \citet{koristagoad}'s calculations are all for a hard active galactic nucleus SED including a luminous X-ray tail\footnote{The command in Cloudy is ``agn 6.683, -1.2, -1.2, -0.9'' (K. Korista, personal communication), where the numbers in order indicate the logarithm of the temperature of the ``Big Bump'' component, the X-ray to UV ratio $\alpha_{\rm ox}$, the low-energy slope of the Big Bump continuum $\alpha_{\rm UV}$, and the slope of the X-ray component $\alpha_{\rm X}$.}.  Guillochon et al.'s assertion that PS1-10jh had solar composition rests on \citet{koristagoad}'s calculations using this incident SED, but the observed SED in PS1-10jh is quite different from that of an AGN.  
PS1-10jh was followed up by Chandra X-ray Observatory 315 days after peak, and showed an upper limit of only $L_X (0.2 - 10)\keV < 5.8 \times 10^{41} \erg \s^{-1}$.  Incident X-ray power is important because X-rays heat the gas, which leads to reduced recombination coefficients \citep[e.g.,][]{vernerferland} and also smaller optical depths in lines.
Gezari et al. fit the observed SED with single blackbody curves at effective temperatures of $T_{\rm eff,low} = 2.9\times 10^4\K$ and $T_{\rm eff,high}=5.5\times 10^4\K$ (depending on the unknown level of extinction).  

We perform Cloudy calculations for four incident SED shapes, and a range of radiation fluxes and densities.  For each model, we perform calculations using $n_{\rm res} =10$, 15, 25, and 35 ``resolved'' atomic energy levels for hydrogen and helium, to check convergence:  when a calculation is converged, increasing the number of resolved energy levels should not change the results.  We calculate the ratio of output fluxes in an individual line between the $n_{\rm res}=25$ and 15 calculations (``Ratio I'' for H$\alpha$ and He4686), and between the $n_{\rm res}=35$ and 25 calculations (``Ratio II'' for H$\alpha$ and He4686).  In some cases (especially at high density and low incident flux), the calculations for $n_{\rm res}=35$, and occasionally for $n_{\rm res}=25$, did not complete in 48 hours (on an Intel Xeon 1.6 GHz processor).  We adopt as our criterion of convergence that (1) at least $n_{\rm res}=25$ must have completed successfully, and (2) if $n_{\rm res}=35$ completed successfully, then Ratio II must lie between 0.9 and 1.1 for both H$\alpha$ and He4686; if only $n_{\rm res}=25$ completed, then Ratio I must lie between 0.9 and 1.1 for both H$\alpha$ and He4686.  Otherwise, we consider the calculation not to have converged.  For converged calculations, we report values from $n_{\rm res}=35$ if available, or from $n_{\rm res}=25$ if not. 

\begin{figure*}
\centerline{\epsfig{file=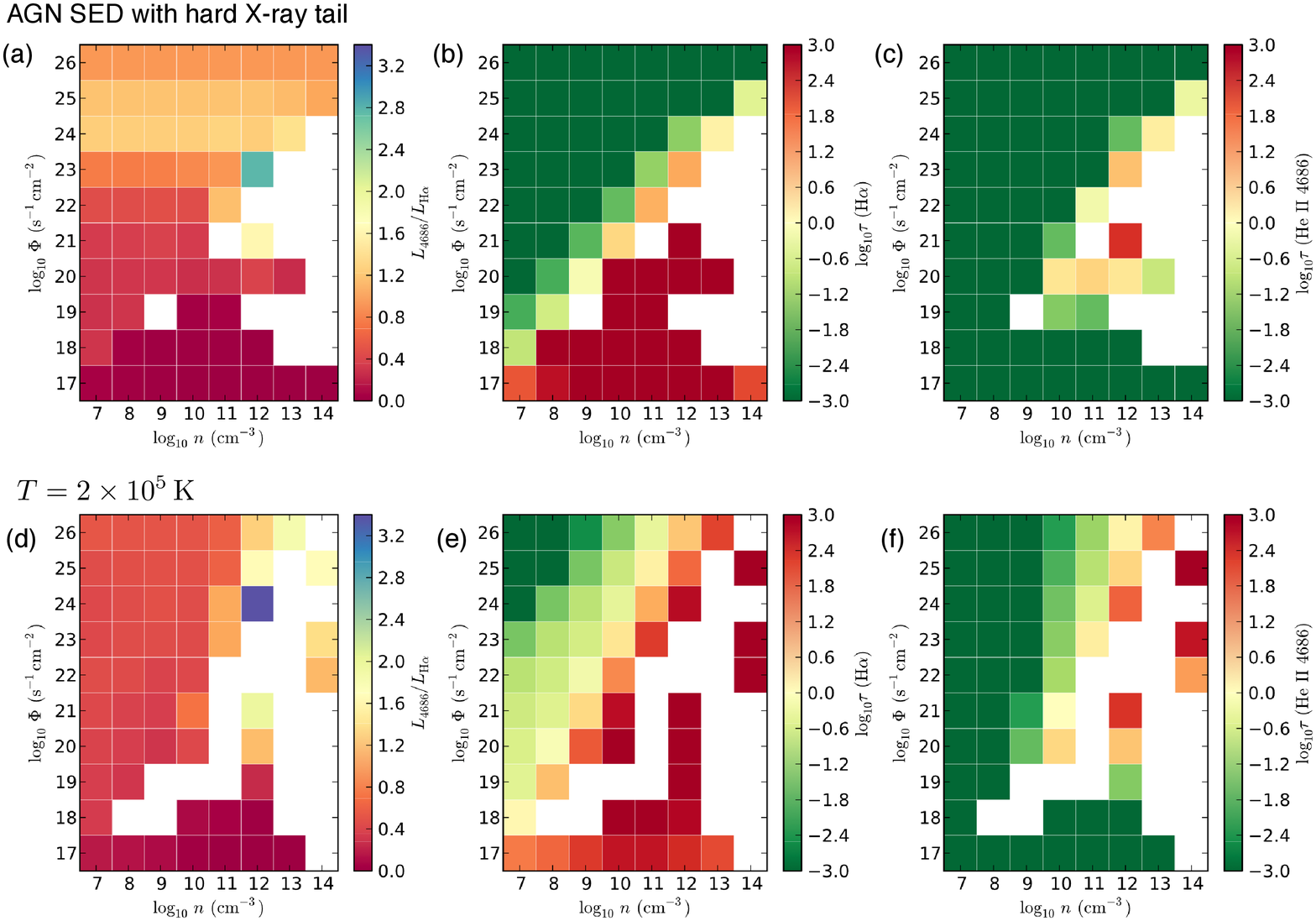, width=7in}}
\caption{Panel (a):  Line ratio $\LHeHa$ for a range of incident fluxes and number densities calculated with Cloudy C13.01, similar to \citet{koristagoad}'s Figure 1 calculated with Cloudy v90.04: an AGN incident spectral energy distribution (see text), a column density of $N_{\rm H} = 10^{23}\cm^{-2}$, and cosmic abundances.  White squares represent models that did not converge.  The maximum calculated line ratio is 2.8, almost a factor of 2 below the lower limit observed by \citet{gezari12} for PS1-10jh.  Panels (b) and (c):  Optical depths in the lines H$\alpha$ (b) and He II 4686 (c), for the same Cloudy calculation as in (a).  The highest line ratio $\LHeHa$ comes from the region where H$\alpha$ optical depths are high but He II 4686 less so, so that H$\alpha$ is relatively suppressed.  Panels (d) - (f): Same as Panels (a) - (c) but for a thermal incident SED with temperature $T = 2\times 10^5\K$. The maximum calculated line ratio is 3.4.  Qualitative similarities and differences between results for different SEDs are described in the text.  \label{fig:AGN_GMR_lineratio_linetau}}
\end{figure*}

\begin{figure*}
\centerline{\epsfig{file=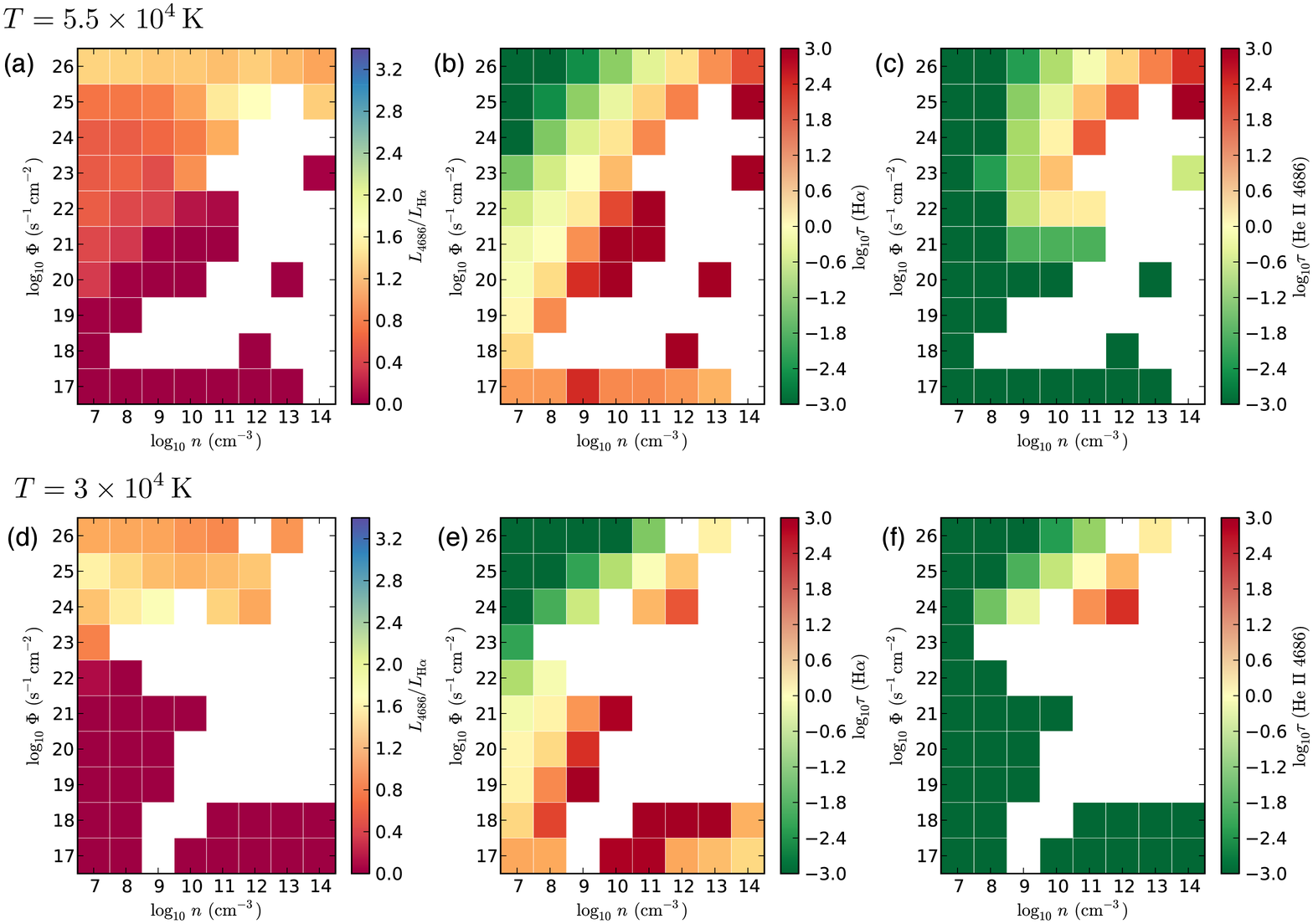, width=7in}}
\caption{Same as Figure \ref{fig:AGN_GMR_lineratio_linetau}, but for thermal incident SEDs having temperatures $T = 5.5\times 10^4\K$ (Panels (a) - (c)) and $3\times 10^4\K$ (Panels (d) - (f)), in line with the SED observed for PS1-10jh.  The maximum calculated line ratios $\LHeHa$ are 1.7 and 1.8, for $T = 5.5\times 10^4\K$ and $3\times 10^4\K$ respectively, well below the observed lower limit for PS1-10jh.  However, cooler temperatures lead to higher optical depths and more convergence issues for Cloudy, including the region of high flux and high density where largest values of $\LHeHa$ are most likely.  \label{fig:BB3_5p5_lineratio_linetau}}
\end{figure*}

We first recalculate \citet{koristagoad}'s Figure 1 (on which Guillochon et al. base their claim of solar composition).  We perform calculations on a $10\times 8$ logarithmically-distributed grid of density and incident flux, with $n_{\rm H} \in [10^7,10^{14}]\cm^{-3}$ and $\Phi \in [10^{17},10^{26}]\s^{-1}\cm^{-2}$; as \citet{koristagoad} do, we assume cosmic abundances and a column density of $N_{\rm H} = 10^{23}\cm^{-2}$. When highly ionized, this column will have an optical depth to Thomson scattering $\tau_{\rm es} \sim 0.1$, safely below the order-unity limit  for which Cloudy is designed.  We present our results in Panel (a) of Figure \ref{fig:AGN_GMR_lineratio_linetau}.  Each colored square\footnote{Note that Figures \ref{fig:AGN_GMR_lineratio_linetau} and \ref{fig:BB3_5p5_lineratio_linetau} appear with different color tables in the published MNRAS version, at the request of the editor for better black-and-white printing. The data represented in the figures here and those in the published version are the same.} represents the line ratio $\LHeHa$ calculated with Cloudy; white squares represent regions of parameter space where the calculations did not converge (as described above).
For this AGN incident SED, we find that the highest value of $\LHeHa$ is only 2.8 (found for $n_{\rm H}=10^{12}\cm^{-3}$, $\Phi=10^{23}\s^{-1}\cm^{-2}$).

Next we perform Cloudy calculations for thermal SEDs at $T = 2\times 10^5\K$, $5.5\times 10^4\K$ and $3 \times 10^4\K$, at the same range of radiation fluxes and densities as the AGN calculation above. 
 We present our results in Figure \ref{fig:AGN_GMR_lineratio_linetau} Panel (d), and Figure \ref{fig:BB3_5p5_lineratio_linetau} Panels (a) and (d), and find that the highest value of $\LHeHa$ is 3.4 (found for $T = 2\times 10^5\K$, $n_{\rm H}=10^{12}\cm^{-3}$, $\Phi=10^{24}\s^{-1}\cm^{-2}$).

All of these results suggest that large densities and large ionizing fluxes are the most likely conditions to produce significant line ratios $\LHeHa$ (as explained more in \S\ref{sec:cloudy_interp}), but also that the observed line ratio lower limit of 5 is too large to have been achieved with cosmic abundances for either an AGN or a thermal SED (of appropriate temperature).

Observations of the PS1-10jh continuum can indicate the relevant region of ($n$, $\Phi$) parameter space at the time of the first spectrum.
As described in \S\ref{sec:geometry} and Footnote \ref{ft:nspect}, if we assume that the gas has cosmic abundances, we infer that the number density in the shell at that time was $n_{\rm spect} \sim 1 \times 10^{11}\cm^{-3}$ for $\teff = 5.5 \times 10^4\K$, or $n_{\rm spect}\sim 8\times 10^{10}\cm^{-3}$ for $\teff = 3\times 10^4\K$.  For an incident SED with temperature $\teff = 5.5\times 10^4\K$, the hydrogen-ionizing number flux is $\Phi = 1 \times 10^{25}\s^{-1}\cm^{-2}$, and for $\teff = 3\times 10^4\K$, the flux is $\Phi = 4\times 10^{23}\s^{-1}\cm^{-2}$.  Based on these estimates, the region\footnote{Note that \citet{guillochon14} identify a somewhat similar region of parameter space for the time of the first spectrum, $n_{\rm H} \sim 10^{12}- 10^{13}\cm^{-3}$ and $\Phi \sim 10^{23} - 10^{24}\s^{-1}\cm^{-2}$. \citet{gaskell14} also identify $n_{\rm H} \sim 10^{11}\cm^{-3}$ as leading to the largest line ratios, although they do not specify the corresponding flux.} in the suites of Cloudy calculations most relevant to the observations of PS1-10jh is thus $n_{\rm H} \sim 10^{11}\cm^{-3}$ and $\Phi \sim 10^{25}\s^{-1}\cm^{-2}$.  For these parameters, our Cloudy calculations (which are all converged) find line ratios far below the observed lower limit of 5:  $\LHeHa = 1.1, 1.5, 0.6, 1.2$ for thermal SEDs with $T = 3\times 10^4\K, 5.5\times 10^4\K, 2\times 10^5\K$, and the AGN SED.

\subsection{Interpretation of Cloudy results}
\label{sec:cloudy_interp}


Understanding the physical processes important in the different Cloudy calculations will help us decide how to apply them to understanding PS1-10jh.
We begin by explaining the results for a thermal incident SED at $T = 2\times 10^5\K$ depicted in Panels (d) - (f) in Figure \ref{fig:AGN_GMR_lineratio_linetau}, and then explain how the results differ for different incident SEDs.  We examine the different regions of $(n, \Phi)$ in turn.

The top left corner of these plots, where the density is low and the incident ionizing flux is high, corresponds to the simple conditions described around equation (\ref{eq:L4686Ha_simple}) in \S\ref{sec:photoion_review}:  the large ionizing flux ensures that hydrogen and helium both are almost completely ionized throughout the region; meanwhile, the low density keeps the recombination rate low, so the populations of neutral hydrogen and singly ionized helium are tiny, making the region optically thin to both the H$\alpha$ and He II 4686 lines (Panels (e) and (f)).  The line ratio (Panel (d)) is thus close to the simple theoretical value of 0.4 described in equation (\ref{eq:L4686Ha_simple}).

In the bottom part of these plots, the ionizing flux is too small to ionize most of the cloud\footnote{This situation is sometimes called ``radiation bounded;'' its opposite is called ``matter bounded.''}. The majority of hydrogen and helium is neutral, with little H$^{+}$ around to recombine and produce H$\alpha$, but even less He$^{++}$ around to recombine and produce He II 4686 emission, since the ionization potential for helium is higher than for hydrogen (and helium requires two ionizations to become He$^{++}$ rather than one, as hydrogen does to become H$^{+}$). This explains the tiny value for $\LHeHa$ in the bottom part of the figure.  (The bottom part of Panels (e) and (f) shows that H$\alpha$ is optically thick but He II 4686 is optically thin---because He II absorption would require a substantial population of He$^+$ ions, but almost all of the helium is neutral.)

In the top right part of the plots, the ionizing flux is large enough to keep the region almost completely ionized, yet the density is large enough that a relatively high recombination rate maintains small but significant populations of neutral hydrogen and singly ionized helium.  Under these conditions, the H$\alpha$ and $4686\ang$ lines can become optically thick, and processes such as collisions and 3-body recombination become important.  The limiting behavior, as the cloud becomes optically thick at all wavelengths, is that the line ratio tends towards the ratio of blackbody flux at the line wavelengths, i.e.,
\begin{equation}\label{eq:BB_line_ratio}
\frac{L_{4686}}{L_{\rm H\alpha}} \approx \left(\frac{4686\ang}{6563\ang}\right)^{-4} = 3.8 \, ,
\end{equation}
if these wavelengths are on the Rayleigh-Jeans tail (where $\nu B_\nu \propto \nu^4$). As the cloud becomes more and more optically thick, the lines become undetectable against the bright continuum.

Importantly for us, there can be an intermediate regime where the optical depth to H$\alpha$ is large while the optical depth to He II 4686 is not so large, and so the H$\alpha$ emission can be somewhat suppressed relative to the He II 4686 emission (see also \citealt{gaskell14}).  This is the regime where our calculations generally find the values of $\LHeHa$ to be largest, and also where our earlier estimates based on the continuum suggest PS1-10jh was at the time of the first spectrum (see \S\ref{sec:cloudy_calcs}).  We emphasize that unfortunately, these high fluxes and optical depths also lead to convergence issues for Cloudy, making Cloudy an ineffective tool for exploring this key region of parameter space.

Now we compare these results for $T = 2 \times 10^5\K$ with the results for other incident SEDs.  Looking ahead to the other thermal SEDs shown in Figure \ref{fig:BB3_5p5_lineratio_linetau}, we see that the structure is very similar, with slightly different temperatures leading to slightly different optical depths and line ratios.  The maximum line ratios $\LHeHa$ are only 1.7 and 1.8 for $5.5\times 10^4\K$ and $3\times 10^4\K$ respectively---but much of the region at large $\Phi$ and large $n$ is unconverged, so perhaps we are missing larger line ratios there.

Back to Figure \ref{fig:AGN_GMR_lineratio_linetau}, the structure seen in Panels (a) - (c) in is similar to that in Panels (d) - (f), but the line ratio in the high-flux region of the figure is close to 1 rather than 0.4.  The reason is that the AGN incident SED contains a luminous X-ray tail; for the same incident number flux $\Phi$, the hard AGN SED is more energetic than the thermal SED, and so the gas is much hotter (by a factor of $\sim 100$ or more).  Recombination coefficients depend on temperature:  at low temperatures they scale roughly as $\alpha \propto T^{-0.5}$, while at high temperatures they scale as $\alpha \propto T^{-1.5}$, with the dividing line around $T/Z^2 \sim 10^6\K$ \citep{ferland92, vernerferland}.  Recombination coefficients for hydrogen-like ions are related by $\alpha(Z,T) \approx Z\alpha(1,T/Z^2)$ \citep[e.g.,][]{osterbrock, vernerferland}, and thus at high temperatures, $\alpha_{4686}$ is higher relative to $\alpha_{\rm H\alpha}$ than at low temperatures. This in equation (\ref{eq:L4686Ha_simple}) explains the line ratios in the top part of Panel (a) in Figure \ref{fig:AGN_GMR_lineratio_linetau}.

We also note that for the thermal SEDs at the highest values of $\Phi$ and $n$, the neutral fraction of hydrogen is high enough that the gas becomes optically thick to H$\alpha$, and so the line ratio $\LHeHa$ can rise above unity. By contrast, the AGN SED heats the gas so much that collisional ionization is important as well as photoionization, and so the neutral fraction of hydrogen remains low, the gas remains optically thin to H$\alpha$, and the line ratio $\LHeHa$ remains close to 1.

Because of the qualitatively different behaviors seen in the top row relative to the bottom row of plots in Figure \ref{fig:AGN_GMR_lineratio_linetau}, we disagree with \citet{gaskell14}'s statement that results do not depend on ionization parameter ($U \equiv \Phi/n c$) or SED shape: Rather, it is crucial to perform calculations using the appropriate incident SED in order to properly capture how the temperature of the gas affects optical depths and line emission.  No X-ray observations of PS1-10jh were taken around the time of the spectrum, so we cannot know the true shape of the incident SED that produced it; however, the late-time Chandra observation showed no X-rays.  Further investigations of photoionization in tidal disruption events must go beyond using only AGN SEDs.

Finally, we point out that workers in the field occasionally offer the explanation that H$\alpha$ is suppressed relative to He II 4686 because the hydrogen is ``over-ionized.''  We emphasize that H$\alpha$ emission is produced following ionized hydrogen H$^{+}$ recombining with an electron (and He II 4686 is produced by He$^{++}$ recombining with an electron):  a highly ionized gas cloud produces {\it more} H$\alpha$ emission than a partially ionized one, not less.  In optically thin systems, the number density of neutral hydrogen is irrelevant to determining the luminosity of H$\alpha$ emission (see \S\ref{eq:L4686Ha_simple}); H$^0$ becomes important in this story only for absorbing H$\alpha$ photons and suppressing H$\alpha$ emission.

\subsection{Problems with applying Cloudy calculations to PS1-10jh}
\label{sec:cloudy_problems}
Having understood the behavior of the line ratio $\LHeHa$ in these suites of Cloudy calculations, we are positioned to ask how applicable these results are to PS1-10jh, and what they can tell us about the nature of the event.

A serious problem in applying these results to PS1-10jh is that the Cloudy calculations assume that the gas is {\it stationary}, while the observed linewidth of He II 4686 $\sim 9000\km\s^{-1}$ assures us that it is not.  The lines in Cloudy are only thermally broadened, by $\sim 10\km\s^{-1}$ for typical gas temperatures of $\sim {\rm few}\times 10^4 - 10^5\K$ (appropriate for the thermal incident SEDs), or up to a few $\times 100\km\s^{-1}$ for the most extreme cases with the hard AGN incident SED.  This means that the line broadening should typically be $\sim 10^2 - 10^3$ times greater than in the calculations.  Crucially, this should diminish the optical depths in H$\alpha$ and He II 4686 by this same factor of $10^2 - 10^3$:  Doppler shifting due to large velocity gradients in the gas allows line photons to escape that would otherwise be trapped.  We thus expect the gas to be close to optically thin to H$\alpha$ (and He II 4686) even in the high flux, high density cases where our (stationary) Cloudy calculations showed H$\alpha$ to be optically thick (see Panels (b) and (e) in Figures \ref{fig:AGN_GMR_lineratio_linetau} and \ref{fig:BB3_5p5_lineratio_linetau})---and where we found the highest values of $\LHeHa$.  This should alter the radiative transfer \citep{koristagoad2000}, and likely {\it reduce} $\LHeHa$ towards its optically thin value ($\sim 0.4$ or $\sim 1$, depending on the gas temperature), even further away from the observed lower limit for PS1-10jh\footnote{An additional concern is that broadening lines makes them appear fainter relative to the continuum; to agree with the observations, photoionization calculations must result in a He II 4686 line that is broad and nevertheless significantly brighter than the continuum.}.  We believe this point casts the most serious doubt on \citet{guillochon14} and \citet{gaskell14}'s claim that PS1-10jh was produced from gas of cosmic abundances and could thus have been the disruption of a main sequence star.

Another concern is that Cloudy requires conditions that are optically thin to Thomson scattering, $\tau_{\rm es} \lesssim 1$.  As argued in \S\ref{sec:rad_photo}, the observed blackbody continuum indicates that the gas giving rise to the emission is optically thick, and Thomson scattering is the dominant opacity for the appropriate temperature and density regime.  We perform our Cloudy calculations by dividing the gas into two pieces: the Thomson-thin outer piece that we simulate with Cloudy, and the Thomson-thick inner piece whose continuum emission (which we simulate simply by choosing various incident SEDs) irradiates the Thomson-thin outer piece.  To investigate spectra of TDEs properly, one should simulate the full extent of the gas all at once, including physics describing the luminosity generation (presumably) deep inside, and Thomson scattering. Unfortunately, Cloudy is not an appropriate tool for such studies.

\subsection{Conclusions and directions for future work}\label{sec:cloudy_conclusions}

\citet{gezari12} measured a lower limit of $\LHeHa > 5$ in the spectrum of PS1-10jh, taken 22 days before peak luminosity.  In this appendix, we have presented Cloudy calculations aimed to answer the question of whether this observed line ratio could have been produced by gas of cosmic abundance; if not, PS1-10jh presumably could not have been the disruption of a solar-type star, as has been claimed by \citet{guillochon14} and \citet{gaskell14}.

We used Cloudy to calculate line ratios for a suite of gas densities and incident ionizing fluxes, for a hard X-ray-bright SED and three thermal SEDs, and cosmic abundances.  From these, we draw the following conclusions:
\begin{enumerate}
\item Results can be qualitatively different (see upper and lower rows of plots in Figure \ref{fig:AGN_GMR_lineratio_linetau}) depending on the shape of the incident SED, as X-rays heat the gas to high temperatures and reduce the ratio of He$^{++}$ recombination relative to H$^+$ recombination.  Previous work has used hard AGN SEDs, while the observations indicate a thermal SED at $\teff \sim 3\times 10^4\K - 5.5 \times 10^4\K$ with no evidence for X-rays.
\item The highest line ratios we calculate are $\LHeHa = 2.8$ for the AGN SED and 3.4 for the thermal SED at $2\times 10^5\K$.  These both lie below the observed lower limit of 5 found by \citet{gezari12}, and close to or below the lower limit of $3.7\pm 25\%$ inferred by \citet{gaskell14}.
\item The highest line ratios are found for models having large ionizing flux $\Phi$ and large density $n$: the gas is highly ionized throughout, but the high density leads to a high recombination rate and so a relatively large neutral fraction of hydrogen, and so a large optical depth to H$\alpha$. This can suppress H$\alpha$ emission relative to He II 4686.
\item However, the Cloudy calculations are performed for stationary gas. Large velocity gradients in the gas, as indicated by the observed He II 4686 linewidth $\sim 9000\km\s^{-1}$, reduce line optical depths dramatically, and could reduce the line ratio $\LHeHa$ from $\sim 3$ at maximum down to $\sim 0.4 - 1$, well below the observed line ratio.
\item Cloudy has convergence issues as the number of ``resolved'' hydrogen levels increases, especially at high fluxes and densities.  We suggest that these issues likely affect the validity of Cloudy results in \citet{koristagoad} and \citet{gaskell14} as well.
\end{enumerate}

We thus assert that, although these calculations are not conclusive, it is unlikely that the spectrum of PS1-10jh was produced by gas of cosmic abundance, and thus it is unlikely that PS1-10jh was the disruption of a solar-type star.  For these reasons, this paper investigates instead the possibility that PS1-10jh was the disruption of a stripped helium core.

For future theoretical studies of tidal disruption spectra, we suggest Monte Carlo radiative transfer studies, such as those currently being undertaken by N. Roth et al. (personal communication).  Important questions to examine include:
\begin{enumerate}
\item How do large velocity gradients affect the spectrum?
\item How does optically thick Thomson scattering affect the spectrum?
\item If a predicted line ratio $\LHeHa$ is large but the gas is optically thick at many wavelengths, are the lines bright enough to measure against the bright continuum background?  Can theoretical calculations reproduce not only line ratios but also line intensities relative to the continuum?
\item How do results vary for different abundance ratios of helium to hydrogen?  How much must hydrogen be depleted to produce the observed spectrum of PS1-10jh?
\end{enumerate}

\section{He II lines, continued:  ionization equilibrium and optical depth}
\label{app:heII_lines}

In this section, we justify analytically several assumptions about He II lines made in \S\ref{sec:heII_emission}.

\subsection{Helium ionization and equilibrium}
\label{sec:he_equilib}
We solve the Saha equation to estimate the ionization state of helium.  Assuming that the gas is purely helium, the ratio of doubly ionized to singly ionized helium is 
\begin{equation}
\frac{\nhepp}{n_{\rm He^+}} = \frac{1}{n_{\rm e}}\left(\frac{2\pi m_{\rm e}kT}{h^2}\right)^{3/2}\frac{2U_{\rm He^{++}}}{U_{\rm He^+}}e^{-\chi_{\rm He^+}/kT} \, ,
\end{equation}
where $U_{\rm He^{++}} \approx 1$ and $U_{\rm He^+} \approx 2$ are the partition functions, and $\chi_{\rm He^+} = 54.4\eV$ is the ionization potential from He$^+$ to He$^{++}$.  In this section, our fiducial value for the number density in the shell at the time of the first spectrum will be $n_{\rm spect} \sim 4\times 10^{10}\cm^{-3}$ (see eq. \ref{eq:nspect}).  We find that for a wide range in densities of pure helium around this value, $n_{\rm He} \sim 10^8\cm^{-3} - 10^{12}\cm^{-3}$,
the transition from almost completely singly-ionized to almost completely doubly-ionized takes place steeply over a narrow range of temperature $T \approx (2-3)\times 10^4\K$.  From photometric measurements, the temperature at the photosphere is inferred to be at least $3\times 10^4\K$ \citep{gezari12}.  Although the gas outside the photosphere at $\rph$ may not have exactly this temperature (since the gas is optically thin in the continuum), its temperature likely is not too different from this since the gas remains coupled to the radiation through line transitions.  Therefore we expect that the gas is almost completely doubly ionized.  For example, the fraction of singly ionized helium is $n_{\rm He^+}/\nhepp \approx 9 \times 10^{-3}$  
at $T = 3\times 10^4\K$ and $\nhepp = 4\times 10^{10}\cm^{-3}$, our estimates for the conditions in the shell at the time of the first spectrum.  Furthermore, almost all of the singly ionized helium is in the ground state, since $kT \ll \chi_{\rm He^+}$.

The time for the gas to reach this ionization equilibrium is the recombination time,
\begin{eqnarray}\label{eq:trec}
t_{\rm rec} & \sim & (n_{\rm e}\alpha_{\rm He^{++}})^{-1} \\
& \sim & 10\left(\frac{n_{\rm e}}{8\times 10^{10}\cm^{-3}}\right)^{-1} \s \, ,
\end{eqnarray}
%
where $\alpha_{\rm He^{++}}\sim 8\times 10^{-13}\cm^3\s^{-1}$ is the Case B He$^{++}$ recombination coefficient \citep[p.38]{osterbrock} at $T=4\times 10^4\K$ (which grows weakly with decreasing temperature).
This timescale is much shorter than the expansion timescale of the shell ($\sim 70\days$ to reach peak) or timescales of observations (days).  The gas therefore is easily in ionization equilibrium, and should be expected to produce He II emission line radiation.

\subsection{Optical depth to He II photons}
\label{sec:tau_heII}

Here we show that the region just outside $\rph$ is highly optically thick to helium Lyman $\alpha$ photons at $304\ang$ (He$^{+}$: $n=1\rightarrow 2$), and therefore capable of supporting a line-driven wind.  We also show that this region is marginally optically thin to He II 4686$\ang$ photons (He$^{+}$: $n=3\rightarrow 4$), allowing us to see these photons as a broad emission line.  We estimate optical depths in these lines by comparing with the (local) optical depth to Thomson scattering, $d\tau_{\rm es}$.

The optical depth to $\lambda_{\rm He L\alpha}=304\ang$ photons along a differential path length is
\begin{equation}
d\tau_{\rm He L\alpha} = \frac{n_{\rm He^+}}{\nhepp}\frac{n_{\rm He^+,1}}{n_{\rm He^+}}\frac{\sigma_{\rm He L\alpha}}{\sigma_{\rm T}}\,d\tau_{\rm es} \, ,
\end{equation}
where $n_{\rm He^+,1}$ is the number density of $n_{\rm He^+}$ ions in the ground state, and $\sigma_{\rm He L\alpha}$ is the cross section to He L$\alpha$ photons. The thermally-broadened cross section to this transition is
\begin{eqnarray}
\label{eq:sigma_helath}
\sigma_{\rm He L\alpha, th} & = & \frac{3\lambda_{\rm He L\alpha}^3}{8\pi}\left(\frac{4m_{\rm p}}{2\pi kT}\right)^{1/2}A_{\rm He,21} \\
& \approx & 2\times 10^{-14}\left(\frac{T}{3\times 10^4\K}\right)^{-1/2}\cm^2 \approx 3\times 10^{10}\sigma_{\rm T} \nonumber
\end{eqnarray}
%
\citep[p.77]{osterbrock} with the Einstein $A$ coefficient\footnote{\label{ft:einsteina}Einstein $A$ coefficients come from ``Persistent Lines of Singly Ionized Helium (He II),'' National Institute of Standards and Technology Physical Meas. Laboratory, http://physics.nist.gov/PhysRefData/Handbook/Tables/heliumtable4.htm.  When we use these values in equations (\ref{eq:sigma_helath}) and (\ref{eq:sigma_4686}), we account for the different fractions of ions in the different spin states according to their degeneracies.}  
$A_{\rm He,21} \approx 1\times 10^{10}\s^{-1}$.
To calculate the full optical depth along a single line of sight, we must account for the fact that photons pass through gas at a variety of velocities whose spread ($\Delta v \sim 4500\km\s^{-1}$, the observed half linewidth of the $4686\ang$ emission line) is much larger than the thermal velocity ($\sim 20\km\s^{-1}$ for $T \sim 3\times 10^4\K$).  Calculating this in detail involves determining the velocity profile as a function of radius, and integrating along different lines of sight through the expanding gas. However, as we don't know the velocity profile well enough, we instead simply approximate that absorption is spread out evenly between wavelengths $\lambda_{\rm He L\alpha}(1-\Delta v/c)$ and $\lambda_{\rm He L\alpha}$, 
so the velocity-broadened cross section is 
\begin{eqnarray}
\sigma_{\rm He L\alpha, v} & = & \sigma_{\rm He L\alpha, th}\left(\frac{v_{\rm th}}{\Delta v}\right) \\
& \approx & 10^{8} \left(\frac{v}{4500\km\s^{-1}}\right)^{-1}\sigma_{\rm T} \, .
\end{eqnarray}
%
From \S\ref{sec:he_equilib}, the singly ionized fraction is $n_{\rm He^+}/\nhepp \sim 9\times 10^{-3}$ for $T = 3\times 10^4\K$ and $\nhepp = 4\times 10^{10}\cm^{-3}$, and essentially all He$^+$ ions are in the ground state; the optical depth to He L$\alpha$ photons is thus
\begin{equation}
d\tau_{\rm He L\alpha} \sim 10^6 \left(\frac{n_{\rm He^+}/\nhepp}{9\times 10^{-3}}\right)\left(\frac{\Delta v}{4500\km\s^{-1}}\right)^{-1}d\tau_{\rm es} \, :
\end{equation}
the gas outside the electron scattering photosphere (where $\tau_{\rm es}\lesssim 1$) is highly optically thick to He L$\alpha$, and therefore radiation pressure on this line transition can drive a wind.

A similar calculation shows that the region is optically {\it thin} to He II $4686\ang$ photons.
The local optical depth to these photons is
\begin{equation}
d\tau_{4686} = \frac{n_{\rm He^+}}{\nhepp}\frac{n_{\rm He^+,3}}{n_{\rm He^+}}\frac{\sigma_{4686}}{\sigma_{\rm T}}\,d\tau_{\rm es} \, ,
\end{equation}
with velocity-broadened cross section
\begin{eqnarray}
\label{eq:sigma_4686}
\sigma_{4686,{\rm v}} & = & \frac{3\lambda_{4686}^3}{8\pi}v^{-1}A_{\rm He,43} \\
& \approx & 7\times 10^{9} \left(\frac{v}{4500\km\s^{-1}}\right)^{-1}\sigma_{\rm T} \, , \nonumber
\end{eqnarray}
and $A_{\rm He,43} \approx 1\times 10^8\s^{-1}$ (see footnote \ref{ft:einsteina}).
For a gas temperature $T = 3\times 10^4\K$, the fraction of He$^+$ ions in the $n=3$ energy level is $n_{\rm He^+,3}/n_{\rm He^+} = 3^2e^{-(1-1/3^2)\chi_{\rm He^+}/kT}\approx 7\times 10^{-8}$, 
far less than the fraction in the ground state.  The optical depth to $4686\ang$ photons is order unity just at the electron scattering photosphere $\rph$, and falls outside,
\begin{equation}
d\tau_{4686} \sim 4 \left(\frac{n_{\rm He^+}/\nhepp}{9\times 10^{-3}}\right)\left(\frac{n_{\rm He^+,3}/n_{\rm He^+}}{7\times 10^{-8}}\right)\left(\frac{\Delta v}{4500\km\s^{-1}}\right)^{-1}d\tau_{\rm es} \, .
\end{equation}
%
That the gas is marginally optically thin to these photons outside the electron scattering photosphere justifies our approximation of using the observed line luminosity to estimate the emission measure $(EM)$ in \S\ref{sec:emission_measure}.

\subsection{Line-driven wind mass-loss rate}
\label{sec:mdotwind}
The rate of mass loss due to the line-driven wind $\dot{M}_{\rm wind}$ can be estimated by setting the rate of momentum imparted by photons to the gas to the rate of momentum carried by the outflowing gas.  Since the majority of He$^{+}$ ions are in the ground state (\S\ref{sec:he_equilib}), photons need to have an energy greater than or equal to the He Ly$\alpha$ transition in order to exert pressure on the helium ions.  We estimate the mass-loss rate
\begin{eqnarray}
\dot{M}_{\rm wind} & \sim & \frac{L(h\nu > 40.8\eV)}{c v_\infty} \\
& \sim & 0.02 \left(\frac{L_{\rm >40.8}}{2\times 10^{43}\erg/\s}\right)\left(\frac{v_\infty}{4500\km/\s}\right)^{-1} \msun\yr^{-1}\nonumber \, ,
\end{eqnarray}
where we have normalized the velocity to that of the observed half linewidth of the $4686\ang$ line, and $L(h\nu > 40.8\eV)$ to a value corresponding to a radiation temperature of $\teff \sim 5.5 \times 10^4\K$ with $\rph\sim 3\times 10^{14}\cm$, appropriate for that temperature at the time of the first spectrum.  However, because $40.8\eV$ is on the exponential tail of the blackbody, this estimate is very sensitive to the assumed shape of the incident SED.  The predicted mass-loss rate would be 3 orders of magnitude lower for a blackbody with $T \sim 3\times 10^4\K$ at the same time of observation, but could be much larger if the event produced significant X-rays (which is possible since no X-ray observations were made until almost a year after the optical discovery).

Building on this very uncertain estimate, the wind's density would be
\begin{eqnarray}
\rho_{\rm wind} & \sim & \frac{\dot{M}_{\rm wind}}{4\pi r^2 v_{\rm wind}} \\
& \sim & 2\times 10^{-15}\left(\frac{r}{\rph}\right)^{-2}  \g\cm^{-3} \nonumber \\
& & \times \left(\frac{\dot{M}_{\rm wind}}{0.02\msun\yr^{-1}}\right)\left(\frac{\rph}{3\times 10^{14}\cm}\right)^{-2}\left(\frac{v_{\rm wind}}{4500\km\s^{-1}}\right)^{-1} \, . \nonumber
\end{eqnarray}
This density is $\sim 10^2$ times smaller than the density we inferred for the shell. However, we have neglected the effects of metal ions, including the lines that are prominent in O-star winds (C IV, Si IV, N V, etc., which are lower energy than He L$\alpha$), so this estimate should be considered a firm lower limit to the wind density.
Spectroscopic observations of ultraviolet resonance lines in absorption (such as helium Lyman $\alpha$, and the UV resonance lines just mentioned) would allow better estimates of the optical depth and mass in the wind.

\section{The photosphere in a scattering-dominated flow}
\label{sec:scat_flow}

Our analysis in \S\ref{sec:rad_photo} assumes that the emission from PS1-10jh is blackbody, as do analogous calculations for supernova observations \citep[e.g.,][]{quimby, chornock14}.  In reality, the dominant opacity in the flow is Thomson scattering rather than true absorption, meaning that the emission in the outermost layer of the flow is ``modified blackbody'' rather than true blackbody.  Our preliminary estimates (using the photoionization code Cloudy) suggest that at peak emission, the absorption opacity is $10^{-3}-10^{-2}$ times the scattering opacity, and larger earlier on (when the density is higher).  The spectral energy distribution of a modified blackbody is broader than that of a true blackbody, so the temperatures inferred from the observations may be (mild) underestimates.

These effects reduce the assumed emission by $(\kappa_{\nu, \rm abs}/\kappa_{\nu, \rm scat})^{1/2}$ (where $\kappa_{\nu, \rm abs}$ and $\kappa_{\nu, \rm scat}$ are the absorption and scattering opacities), though this reduction is slightly counterbalanced by the higher true temperature.  This implies that the radius of the photosphere may actually be a factor of a few larger than we estimated in \S\ref{sec:rad_photo} and Figure \ref{fig:roft}, implying a larger velocity, larger mass of the shell, and larger kinetic energy.

Specifically, for a photosphere that is larger than our previous estimate by $f \sim 3$, the velocity would be $v_0\sim 3000\km\s^{-1}$, the mass of the shell would be $M_{\rm shell}\sim 0.09\msun$, and the kinetic energy would be $E_{\rm K} \sim 8 \times 10^{47}\erg$.  The kinetic energy would still be far below the energy of emitted radiation.  The shell would carry $\sim 60\%$ rather than $7\%$ of the (originally bound) half of the mass of the disrupted core.

We defer to future work more detailed investigation of the photospheric radius accounting for deviations from blackbody.

\label{lastpage}

\end{document}